\documentclass[preprint,12pt]{elsarticle}
\usepackage{amsmath}
\usepackage{amssymb}
\usepackage{supertabular}
\usepackage{theorem}
\usepackage{yhmath}

\newtheorem{theorem}{Theorem}
\newtheorem{lemma}{Lemma}
\newtheorem{corollary}{Corollary}
\newtheorem{proposition}{Proposition}
\newdefinition{nnote}{Note}
\newproof{pf}{Proof}

\def\vec#1{\mathchoice
{\mbox{\boldmath$\displaystyle#1$}} {\mbox{\boldmath$\textstyle#1$}}
{\mbox{\boldmath$\scriptstyle#1$}}
{\mbox{\boldmath$\scriptscriptstyle#1$}}}

\def\bbbz{{\rm\bf Z}}

\begin{document}

\begin{frontmatter}

\title{$x^{2^l+1}+x+a$ and Related Affine Polynomials over $\mathrm{GF}(2^k)$\tnoteref{NFR}}
\author{Tor Helleseth}
\ead{Tor.Helleseth@uib.no}
\author{Alexander Kholosha}
\ead{Alexander.Kholosha@uib.no}
\address{The Selmer Center, Department of Informatics, University of
Bergen, PB 7800, N-5020 Bergen, Norway}

\tnotetext[NFR]{Research supported by the Norwegian Research
Council.}

\begin{abstract}
In this paper, the polynomials $P_a(x)=x^{2^l+1}+x+a$ with
$a\in\mathrm{GF}(2^k)$ are studied. New criteria for the number of
zeros of $P_a(x)$ in $\mathrm{GF}(2^k)$ are proved. In particular, a
criterion for $P_a(x)$ to have exactly one zero in
$\mathrm{GF}(2^k)$ when $\gcd(l,k)=1$ is formulated in terms of the
values of permutation polynomials introduced by Dobbertin. We also
study the affine polynomial $a^{2^l}x^{2^{2l}}+x^{2^l}+ax+1$ which
is closely related to $P_a(x)$. In many cases, explicit expressions
for calculating zeros of these polynomials are provided.
\end{abstract}


\end{frontmatter}

\section{Introduction}
 \label{sec:int}
Denote $\mathrm{GF}(2^k)$ a finite field with $2^k$ elements, let
$\mathrm{GF}(2^k)^*=\mathrm{GF}(2^k)\setminus\{0\}$ and
$\mathrm{GF}(2^k)^{**}=\mathrm{GF}(2^k)\setminus\{0,1\}$. Take
positive integers $k$ and $l$ with $l<k$. The focus of this paper are
the following polynomials over $\mathrm{GF}(2^k)$:
\[P_a(x)=x^{2^l+1}+x+a\]
with $a\in\mathrm{GF}(2^k)^*$. It is clear that $P_a(x)$ does not
have multiple roots. These polynomials have recently arisen in
several different contexts that include final geometry, constructing
families of difference sets with Singer parameters \cite{Di02} and
finding crosscorrelation between $m$-sequences
\cite{DoFeHeRo06,HeKhNe07,HeKh08_2}. First, we consider a particular
case when $l$ is coprime to $k$ (which leads to an interesting new
technique based on the use of Dobbertin polynomials) and then we take
a general case with $\gcd(l,k)\geq 1$. With our results, we are able
to distinguish between the case when $P_a(x)$ has none and the case
when it has two zeros in $\mathrm{GF}(2^k)$ if $\gcd(l,k)>1$. This is
considered to be a hard problem in general. Finally, we study the
roots of the following affine polynomial which is shown to be closely
related to $P_a(x)$
\begin{equation}
 \label{eq:F}
F_a(x)=a^{2^l}x^{2^{2l}}+x^{2^l}+ax+1\enspace.
\end{equation}

Polynomials $f(x)=x^{p^l+1}+ax+b$ over a field of characteristic $p$
with an arbitrary $l$ were recently extensively studied by Bluher in
her paper \cite{Bl04}. Thus, here we consider a particular instance
of this problem. However, as a main result of the paper, we prove
new criteria for the number of zeros of $P_a(x)$ in
$\mathrm{GF}(2^k)$. For instance, if $\gcd(l,k)=1$ then the absolute
trace of the particular value of the Dobbertin permutation
polynomial defines whether $P_a(x)$ has a unique zero or not. We
also give explicit polynomial formulas for calculating zeros in case
when zero is unique or there are exactly two of them and $k$ is odd.
Note that $f(x)$ can always be transformed into the form
$x^{p^l+1}+x+c$ by a simple substitution of variable $x$ with $sx$
having $s^{p^l}=a$ (such an $s\in\mathrm{GF}(p^k)$ always exists).
Moreover, even a more general polynomial form
$x^{p^l+1}+ax^{p^l}+bx+c$ can be reduced to $f(x)$ by setting $x$
equal to $x-a$.

In the particular case when $l=1$, the equation $P_a(x)=0$ takes on
the form $D_3(x)=a$ where $D_3(x)=x^3+x$ is the third Dickson
polynomial (a comprehensive reference on this topic is
\cite{LiMuTu93}). Denote
\[\mathcal{H}_i=\{x\in\mathrm{GF}(2^k)^*\ |\ {\rm
Tr}_k(x^{-1})=i\}\quad\mbox{for}\quad i=0,1\] $r_0=\gcd(3,2^k-1)$
and $r_1=\gcd(3,2^k+1)$. Obviously, $r_0=1$ and $r_1=3$ (resp.
$r_0=3$ and $r_1=1$) for $k$ odd (resp. $k$ even). From the
well-known fact (see \cite{LiMuTu93}, \cite[Proposition~5]{DiDo04}
or \cite[Lemma~18]{DoFeHeRo06}) if follows that $D_3(x)$ is a
$r_0$-to-$1$ mapping of $\mathcal{H}_0\setminus\{1\}$ into
$\mathcal{H}_0$ and is a $r_1$-to-$1$ mapping of
$\mathcal{H}_1\setminus\{1\}$ into $\mathcal{H}_1$. Therefore, for
odd $k$ (resp. even $k$) $D_3(x)=a$ has a unique solution in
$\mathrm{GF}(2^k)$ if and only if $a\in\mathcal{H}_0$ (resp.
$a\in\mathcal{H}_1$) which for any $k$ is equivalent to ${\rm
Tr}_k(a^{-1}+1)=1$. In the other cases this equation can have either
none or three solutions.

By the time the earlier version of this paper \cite{HeKh08_1} was
published, we where able to prove many relevant results assuming
additional restrictive conditions. In the current paper, just
Section~\ref{sec:zer1} almost has not been changed compared to
\cite{HeKh08_1}. Sections~\ref{sec:pre}~and~\ref{sec:zer2} had been
considerably revised to contain the results under the the most
general conditions, some proof were rewritten in a simpler way.
Sections~\ref{sec:RelA}~and~\ref{sec:RelL} are completely new. We
believe that a paper containing patches to \cite{HeKh08_1} would be
extremely reader-unfriendly since we would have to refer not just to
the previous results but to the parts of the proofs in
\cite{HeKh08_1}. That is why we decided to submit a self-contained
paper that does not require any prior reading.

\section{Preliminaries}
 \label{sec:pre}
The finite field $\mathrm{GF}(2^d)$ is a subfield of
$\mathrm{GF}(2^k)$ if and only if $d$ divides $k$. The trace and norm
mappings from $\mathrm{GF}(2^k)$ to the subfield $\mathrm{GF}(2^d)$
are defined respectively by
\[{\rm Tr}_d^k(x)=\sum_{i=0}^{k/d-1}x^{2^{id}}\quad\quad\mbox{and}
\quad\quad{\rm N}_d^k(x)=\prod_{i=0}^{k/d-1}x^{2^{id}}\enspace.\] In
the case when $d=1$, we use the notation ${\rm Tr}_k(x)$ instead of
${\rm Tr}_1^k(x)$. In this paper, also let $M_i$ denote the number
of $a\in\mathrm{GF}(2^k)^*$ such that $P_a(x)$ has exactly $i$ zeros
in $\mathrm{GF}(2^k)$.

If $l$ is coprime to $k$, denote $l'=l^{-1}\ (\bmod\;k)$ and recall
the following sequences of polynomials that were introduced by
Dobbertin in \cite{Do99} (see also \cite{DiDo04}):
\begin{eqnarray*}
A_1(x)&=&x\,,\\
A_2(x)&=&x^{2^l+1}\,,\\
A_{i+2}(x)&=&x^{2^{(i+1)l}}A_{i+1}(x)+x^{2^{(i+1)l}-2^{il}}A_i(x)\quad\mbox{for}\quad
i\geq 1\,,\\
B_1(x)&=&0\,,\\
B_2(x)&=&x^{2^l-1}\,,\\
B_{i+2}(x)&=&x^{2^{(i+1)l}}B_{i+1}(x)+x^{2^{(i+1)l}-2^{il}}B_i(x)\quad\mbox{for}\quad
i\geq 1\enspace.
\end{eqnarray*}
These are used to define the polynomial
\begin{equation}
 \label{eq:R}
R(x)=\sum_{i=1}^{l'}A_i(x)+B_{l'}(x)\enspace.
\end{equation}
As noted in \cite{Do99}, the exponents occurring in $A_j(x)$ (resp.
in $B_j(x)$) are precisely those of the form
\[e=\sum_{i=0}^{j-1}(-1)^{\epsilon_i}2^{il}\]
where $\epsilon_i\in\{0,1\}$ satisfy $\epsilon_{j-1}=0$,
$\epsilon_0=0$ (resp. $\epsilon_0=1$), and
$(\epsilon_i,\epsilon_{i-1})\neq(1,1)$.

Further, we will essentially need the following result proven in
\cite[Theorem~5]{Do99} that the polynomial
\begin{equation}
 \label{eq:q(x)}
q^{(\epsilon)}(x)=\frac{\sum_{i=1}^{l'}x^{2^{il}}+\epsilon}{x^{2^l+1}}
\quad\quad\mbox{for}\quad\epsilon=0,1
\end{equation}
is a permutation polynomial on $\mathrm{GF}(2^k)^*$ if and only if
$\epsilon\equiv l'+1\ (\bmod\;2)$. (To be formally more precise, we
get a {\em polynomial} $q^{(\epsilon)}(x)$ if $x^{-(2^l+1)}$ is
substituted by $x^{(2^k-1)-(2^l+1)}$.) In the sequel, we simply use
$q(x)$ instead of $q^{(\epsilon)}(x)$ for $\epsilon\equiv l'+1\
(\bmod\;2)$. Moreover, $q(x)$ and $R(x^{-1})$ are inverses of each
other \cite[Theorem~6]{Do99}, i.e., for any nonzero
$u,v\in\mathrm{GF}(2^k)$ with $q(u)=v^{-1}$ it always holds that
$R(v)=u$. In (\ref{eq:q(x)}) and in the rest of the paper, whenever a
positive integer $e$ is added to an element of $\mathrm{GF}(2^k)$, it
means that added is the identity element of $\mathrm{GF}(2^k)$ times
$e\,(\bmod\;2)$.

In the general case when $\gcd(l,k)=d\geq 1$, let $k=nd$ for some
$n>1$ and introduce a particular sequence of polynomials over
$\mathrm{GF}(2^k)$. For any $u\in\mathrm{GF}(2^k)$ denote
$u_i=u^{2^{il}}$ for $i=0,\dots,n-1$ and let
\begin{eqnarray}
 \label{eq:dC1}
\nonumber C_1(x)&=&1\,,\\
\nonumber C_2(x)&=&1\,,\\
C_{i+2}(x)&=&C_{i+1}(x)+x_i C_i(x)\quad\mbox{for}\quad 1\leq i\leq
n-1\enspace.
\end{eqnarray}

\begin{lemma}
For any $u\in\mathrm{GF}(2^k)$ and $i\in\{1,\dots,n-1\}$
\begin{align}
&C_{i+2}(u)=C_{i+1}^{2^l}(u)+u_1 C_i^{2^{2l}}(u)\quad\quad\mbox{and}\label{eq:dC2}\\
&C_i^{2^l}(u)C_{i+2}(u)+C_{i+1}^{2^l+1}(u)=\prod_{j=1}^i u_j\enspace.\label{eq:C2l}
\end{align}
\end{lemma}

\begin{pf}
Both identities are proved using induction on $i$. For $i=1$ and
$i=2$ the correctness is easily checked taking the definition.
Assuming the identities hold for $i<t$ we get for $i=t>2$
\begin{eqnarray*}
C_{t+2}(u)&\stackrel{(\ref{eq:dC1})}{=}&C_{t+1}(u)+u_t C_t(u)\\
&=&C_t^{2^l}(u)+u_1 C_{t-1}^{2^{2l}}(u)+u_t C_{t-1}^{2^l}(u)+u_t u_1 C_{t-2}^{2^{2l}}(u)\\
&=&(C_t(u)+u_{t-1} C_{t-1}(u))^{2^l}+u_1(C_{t-1}(u)+u_{t-2} C_{t-2}(u))^{2^{2l}}\\
&\stackrel{(\ref{eq:dC1})}{=}&C_{t+1}^{2^l}(u)+u_1
C_t(u)^{2^{2l}}
\end{eqnarray*}
and
\begin{align*}
&C_t^{2^l}(u)C_{t+2}(u)+C_{t+1}^{2^l+1}(u)\\
&\stackrel{(\ref{eq:dC1})}{=}(C_{t-1}^{2^l}(u)+u_{t-1}C_{t-2}^{2^l}(u))(C_{t+1}(u)+u_t C_t(u))+
(C_t(u)+u_{t-1}C_{t-1}(u))^{2^l+1}\\
&=C_{t-1}^{2^l}(u)C_{t+1}(u)+C_t^{2^l+1}(u)+u_{t-1}u_t(C_{t-2}^{2^l}(u)C_t(u)+C_{t-1}^{2^l+1}(u))\\
&\quad{}+u_{t-1}(C_{t-2}^{2^l}(u)C_{t+1}(u)+C_{t-1}(u)C_t^{2^l}(u))\\
&=\prod_{j=1}^{t-1}u_j+\prod_{j=1}^t u_j\\
&\quad{}+u_{t-1}(C_{t-2}^{2^l}(u)(C_t(u)+u_{t-1}C_{t-1}(u))+C_{t-1}(u)(C_{t-1}(u)+u_{t-2}C_{t-2}(u))^{2^l})\\
&=\prod_{j=1}^{t-1}u_j+\prod_{j=1}^t u_j+u_{t-1}(C_{t-2}^{2^l}(u)C_t(u)+C_{t-1}^{2^l+1}(u))=\prod_{j=1}^t u_j\enspace.
\end{align*}
(\ref{eq:dC2}) can be seen as an equivalent recursive definition of
$C_i(x)$.\qed
\end{pf}

We also define polynomials $Z_n(x)$ over $\mathrm{GF}(2^k)$ as
$Z_1(x)=1$ and
\begin{equation}
 \label{eq:Z}
Z_n(x)=C_{n+1}(x)+x C_{n-1}^{2^l}(x)
\end{equation}
for $n>1$. Note that for any $u\in\mathrm{GF}(2^k)$ we get
\begin{eqnarray*}
Z_n^{2^l}(u)&\stackrel{(\ref{eq:Z})}{=}&C_{n+1}^{2^l}(u)+u_1 C_{n-1}^{2^{2l}}(u)\\
&\stackrel{(\ref{eq:dC1})}{=}&C_n^{2^l}(u)+u_0 C_{n-1}^{2^l}(u)+u_1 C_{n-1}^{2^{2l}}(u)\\
&\stackrel{(\ref{eq:dC2})}{=}&C_{n+1}(u)+u_0 C_{n-1}^{2^l}(u)\\
&\stackrel{(\ref{eq:Z})}{=}&Z_n(u)
\end{eqnarray*}
and thus, $Z_n(u)\in\mathrm{GF}(2^l)$. Since
$\mathrm{GF}(2^k)\bigcap\mathrm{GF}(2^l)=\mathrm{GF}(2^d)$, we have
$Z_n(u)\in\mathrm{GF}(2^d)$. The following lemma describes zeros of
$C_n(x)$ in $\mathrm{GF}(2^k)$.

\begin{proposition}
 \label{pr:ZeC}
Take any $v\in\mathrm{GF}(2^{nd})\setminus\mathrm{GF}(2^d)$ with
$n>1$ and let
\begin{equation}
 \label{eq:V}
V=\frac{v_0^{2^{2l}+1}}{(v_0+v_1)^{2^l+1}}\enspace.
\end{equation}
Then
\[C_n(V)=\frac{{\rm
Tr}^{nd}_d(v_0)}{(v_1+v_2)}\prod_{j=2}^{n-1}\left(\frac{v_0}{v_0+v_1}\right)^{2^{jl}}\enspace.\]
If $n$ is odd (resp. $n$ is even) then the total number of distinct
zeros of $C_n(x)$ in $\mathrm{GF}(2^{nd})$ is equal to
$\frac{2^{(n-1)d}-1}{2^{2d}-1}$ (resp.
$\frac{2^{(n-1)d}-2^d}{2^{2d}-1}$). All zeros have the form of
(\ref{eq:V}) with ${\rm Tr}^{nd}_d(v_0)=0$ and occur with
multiplicity $2^l$. Moreover, polynomial $C_n(x)$ splits in
$\mathrm{GF}(2^{nd})$ if and only if $d=l$ or $n<4$.
\end{proposition}

\begin{pf}
First, note that
$\mathrm{GF}(2^{nd})\bigcap\mathrm{GF}(2^l)=\mathrm{GF}(2^d)$ and
$v\in\mathrm{GF}(2^{nd})\setminus\mathrm{GF}(2^d)$ if and only if
$v_0\neq v_1$ which guarantees that the denominator in (\ref{eq:V})
and in the above identity for $C_n(V)$ is not zero. Now, using
induction on $i$ we prove that
\begin{equation}
\label{eq:CiV}
C_i(V)=\frac{\sum_{j=1}^i
v_j}{(v_1+v_2)}\prod_{j=2}^{i-1}\left(\frac{v_0}{v_0+v_1}\right)^{2^{jl}}
\end{equation}
for $2\leq i\leq n+1$. For $i=2$ and $i=3$ this identity is easily
checked using the definition (\ref{eq:dC1}) of $C_i(x)$ (for $i=2$,
we assume the product over the empty set to be equal to $1$).
Assuming this identity holds for $i<t$ we get for $i=t>3$
\begin{align*}
&C_t(V)\stackrel{(\ref{eq:dC1})}{=}C_{t-1}(V)+V_{t-2} C_{t-2}(V)\\
&=\frac{\sum_{j=1}^{t-1}v_j}{(v_1+v_2)}\prod_{j=2}^{t-2}\left(\frac{v_0}{v_0+v_1}\right)^{2^{jl}}+
\frac{v_{t-2}^{2^{2l}+1}\sum_{j=1}^{t-2}v_j}{(v_{t-2}+v_{t-1})^{2^l+1}(v_1+v_2)}
\prod_{j=2}^{t-3}\left(\frac{v_0}{v_0+v_1}\right)^{2^{jl}}\\
&=\frac{\left((v_{t-1}+v_t)\sum_{j=1}^{t-1}v_j+v_t\sum_{j=1}^{t-2}v_j\right)\prod_{j=2}^{t-2}v_0^{2^{jl}}}
{(v_1+v_2)\prod_{j=2}^{t-1}(v_0+v_1)^{2^{jl}}}\\
&=\frac{\sum_{j=1}^t
v_j}{(v_1+v_2)}\prod_{j=2}^{t-1}\left(\frac{v_0}{v_0+v_1}\right)^{2^{jl}}\enspace.
\end{align*}
It remains to note that for $i=n$, in $\mathrm{GF}(2^{nd})$ we have
$\sum_{j=1}^n v_j=\sum_{j=1}^n v^{2^{jd}}={\rm Tr}^{nd}_d(v_0)$.

Obviously, $C_n(V)=0$ if and only if ${\rm Tr}^{nd}_d(v_0)=0$ which
is equivalent to $v_0=u+u^{2^l}$ for some $u\in\mathrm{GF}(2^{nd})$.
This easily follows from the fact that the linear operator
$L(u)=u+u^{2^l}$ on $\mathrm{GF}(2^{nd})$ has the kernel of
dimension $d$ and, thus, the number of elements in the image of $L$
is $2^{d(n-1)}$. For any $u\in\mathrm{GF}(2^{nd})$ we have ${\rm
Tr}^{nd}_d(u+u^{2^l})=0$ leading to the conclusion that the image of
$L$ contains all the elements of $\mathrm{GF}(2^{nd})$ having zero
trace in $\mathrm{GF}(2^d)$ since the total number of such elements
is exactly $2^{d(n-1)}$. Moreover, $u\notin\mathrm{GF}(2^{2d})$
since $v_0\in\mathrm{GF}(2^d)\subseteq\mathrm{GF}(2^l)$ if and only
if the corresponding $u\in\mathrm{GF}(2^{2l})$ and since
$\mathrm{GF}(2^{nd})\cap\mathrm{GF}(2^{2l})=\mathrm{GF}(2^{d\gcd(n,2)})$.
It follows from the proof of Proposition~\ref{pr:P2dZero} that the
mapping from $u\in\mathrm{GF}(2^{nd})\setminus\mathrm{GF}(2^{2d})$
via $v_0=u+u^{2^l}$ to $V\in\mathrm{GF}(2^{nd})^*$ defined by
(\ref{eq:V}) is $(2^{3d}-2^d)$-to-$1$. Therefore, we have found
$\frac{|\mathrm{GF}(2^{nd})\setminus\mathrm{GF}(2^{2d})|}{2^{3d}-2^d}$
distinct zeros of $C_n(x)$ in $\mathrm{GF}(2^{nd})$ and if $n$ is
odd (resp. $n$ is even) then this number is equal to
$\frac{2^{(n-1)d}-1}{2^{2d}-1}$ (resp.
$\frac{2^{(n-1)d}-2^d}{2^{2d}-1}$).

It is easy to check by induction that if $i$ is odd (resp. $i$ is
even) then the algebraic degree of polynomials $C_i(x)$ is equal to
$\frac{2^{il}-2^l}{2^{2l}-1}$ (resp.
$\frac{2^{il}-2^{2l}}{2^{2l}-1}$) since
\[\deg C_{i+2}(x)=\max\{\deg C_{i+1}(x),2^{il}+\deg C_i(x)\}=2^{il}+\deg C_i(x)\enspace.\]
Further, if we define the sequence of polynomials $C'_i(x)$ for
$i=1,\dots,n$ with $C'_1(x)=C'_2(x)=1$ and
$C'_{i+2}(x)=C'_{i+1}(x)+x_{i-1}C'_i(x)$ then $C_i(x)=C'_i(x)^{2^l}$
for $i=1,\dots,n$. Therefore, all zeros of $C_n(x)$ have
multiplicity at least $2^l$. Now it is clear that the number of
zeros having the form of (\ref{eq:V}) with ${\rm Tr}^{nd}_d(v_0)=0$
multiplied by $2^l$ is equal to the degree of $C_n(x)$ if and only
if $d=l$ or $n<4$.

It means that $C_n(x)$ splits in $\mathrm{GF}(2^{nl})$ and zeros of
$C_n(x)$ in $\mathrm{GF}(2^{nd})$ are exactly the elements obtained
by (\ref{eq:V}) using
$w_0\in\mathrm{GF}(2^{nl})\setminus\mathrm{GF}(2^l)$ with ${\rm
Tr}^{nl}_l(w_0)=0$ that result in $V\in\mathrm{GF}(2^{nd})$. It also
follows from the proof of Proposition~\ref{pr:P2dZero} that
polynomial $f_b(y)=y^{2^l+1}+by+b$ with $b\in\mathrm{GF}(2^{nd})^*$
has exactly $2^d+1$ zeros in $\mathrm{GF}(2^{nd})$ if and only if
$b^{-1}$ has the form of (\ref{eq:V}) with ${\rm Tr}^{nd}_d(v_0)=0$.
Take any $V\in\mathrm{GF}(2^{nd})$ obtained by (\ref{eq:V}) using
$w_0\in\mathrm{GF}(2^{nl})\setminus\mathrm{GF}(2^l)$ with ${\rm
Tr}^{nl}_l(w_0)=0$. Then $f_{V^{-1}}(y)$ splits in
$\mathrm{GF}(2^{nl})$ and, by \cite[Corollary~7.2]{Bl04}, this is
equivalent to $f_{V^{-1}}(y)$ having $2^d+1$ zeros in
$\mathrm{GF}(2^{nd})$. Thus, there exists some
$v_0\in\mathrm{GF}(2^{nd})\setminus\mathrm{GF}(2^d)$ with ${\rm
Tr}^{nd}_d(v_0)=0$ that gives this $V$ using (\ref{eq:V}).\qed
\end{pf}

\begin{corollary}
 \label{co:ZeZ}
If $n$ is odd (resp. $n$ is even) then the total number of distinct
zeros of $Z_n(x)$ in $\mathrm{GF}(2^{nd})$ is equal to
$\frac{2^{(n+1)d}-2^{2d}}{2^{2d}-1}$ (resp.
$\frac{2^{(n+1)d}-2^d}{2^{2d}-1}$). All zeros have the form of
(\ref{eq:V}) and occur with multiplicity one. Moreover, polynomial
$Z_n(x)$ splits in $\mathrm{GF}(2^{nd})$ if and only if $d=l$ or
$n=1$.
\end{corollary}

\begin{pf}
Using (\ref{eq:CiV}), it can be verified directly that $C_{n+1}(V)=V
C_{n-1}^{2^l}(V)$ for any $V\in\mathrm{GF}(2^{nd})$ having the form
of (\ref{eq:V}) (the case $n=2$ is easily checked having the
definition of $C_i(x)$). Also, knowing the algebraic degree of
polynomials $C_i(x)$ from the proof of Proposition~\ref{pr:ZeC}, we
conclude that
\[\deg (Z_n(x))=\deg C_{n+1}(x)\]
and is equal to $\frac{2^{(n+1)l}-2^{2l}}{2^{2l}-1}$ (resp.
$\frac{2^{(n+1)l}-2^l}{2^{2l}-1}$) if $n$ is odd (resp. $n$ is
even). Denote
\begin{equation}
 \label{eq:S}
S=\{x\in\mathrm{GF}(2^{nd})\setminus\mathrm{GF}(2^d)\ |\ {\rm
Tr}^{nd}_d(x)\neq 0\}\enspace.
\end{equation}
It follows from the proof of Proposition~\ref{pr:P1Zero} that the
mapping from $v\in S$ to $V\in\mathrm{GF}(2^{nd})^*$ defined by
(\ref{eq:V}) is $(2^d-1)$-to-$1$. Recalling the corresponding fact
from the latest proof, we conclude that the total number of distinct
values of $V$ obtained by (\ref{eq:V}) is equal to
$\frac{|\mathrm{GF}(2^{nd})\setminus\mathrm{GF}(2^{2d})|}{2^{3d}-2^d}+\frac{|S|}{2^d-1}$
being identical to the degree of $Z_n(x)$ if and only if $d=l$ or
$n=1$. Note that two different values of
$v\in\mathrm{GF}(2^{nd})\setminus\mathrm{GF}(2^d)$ with zero and
nonzero trace in $\mathrm{GF}(2^d)$ can not map to the same value
$V$ using (\ref{eq:V}) since $C_n(V)=0$ if and only if the trace of
the corresponding $v$ is also equal zero.

It means that that $Z_n(x)$ splits in $\mathrm{GF}(2^{nl})$ and its
zeros in $\mathrm{GF}(2^{nd})$ are exactly the elements obtained by
(\ref{eq:V}) using
$w_0\in\mathrm{GF}(2^{nl})\setminus\mathrm{GF}(2^l)$ that result in
$V\in\mathrm{GF}(2^{nd})$. It also follows from the proof of
Propositions~\ref{pr:P1Zero} and \ref{pr:P2dZero} that polynomial
$f_b(y)=y^{2^l+1}+by+b$ with $b\in\mathrm{GF}(2^{nd})^*$ has exactly
one or $2^d+1$ zeros in $\mathrm{GF}(2^{nd})$ if and only if
$b^{-1}$ has the form of (\ref{eq:V}). Take any
$V\in\mathrm{GF}(2^{nd})$ obtained by (\ref{eq:V}) using
$w_0\in\mathrm{GF}(2^{nl})\setminus\mathrm{GF}(2^l)$. Then
$f_{V^{-1}}(y)$ has exactly one or $2^l+1$ zeros
$\mathrm{GF}(2^{nl})$ and, by \cite[Corollaries~7.2,~7.3]{Bl04},
this is equivalent to $f_{V^{-1}}(y)$ having one or $2^d+1$ zeros in
$\mathrm{GF}(2^{nd})$ respectively. Thus, there exists some
$v_0\in\mathrm{GF}(2^{nd})\setminus\mathrm{GF}(2^d)$ that gives this
$V$ using (\ref{eq:V}).\qed
\end{pf}

\begin{corollary}
 \label{co:ZeTr}
For any $V\in\mathrm{GF}(2^k)$ having the form of (\ref{eq:V}) with
$n>2$ and ${\rm Tr}^k_d(v_0)\neq 0$ we have ${\rm
Tr}_d^k\left(C_{n-1}^{2^l}(V)/C_n^{2^l+1}(V)\right)=0$.
\end{corollary}

\begin{pf}
Using (\ref{eq:CiV}), it can be verified directly that
\[\frac{C_{n-1}^{2^l}(V)}{C_n^{2^l+1}(V)}={\rm
N}_d^k\left(1+\frac{v_1}{v_0}\right)\frac{v_1\sum_{j=2}^n v_j}{{\rm
Tr}_d^k(v_0)^2}\] for any $V\in\mathrm{GF}(2^k)$ having the form of
(\ref{eq:V}) and $n>2$. Now note that
\[{\rm Tr}_d^k\Big(v_1 \sum_{j=2}^n v_j\Big)={\rm Tr}_d^k\left(v_1
{\rm Tr}_d^k(v_0)+v_1^2\right)={\rm Tr}_d^k(v_0)^2+{\rm
Tr}_d^k(v_0^2)=0\] and we are done.\qed
\end{pf}

Value of $C_i(x)$ is equal to the determinant of a three-diagonal
symmetric matrix (note a comprehensive study of such matrices in
\cite{IlKu85}). Indeed, for any $u\in\mathrm{GF}(2^k)$ and $j\leq i$
let $\Delta_u(j,i)$ denote the determinant of matrix $D$ of size
$i-j+2$ that contains ones on the main diagonal and with
$D(t,t+1)=D(t+1,t)=u_{j+t-1}$ for $t=1,\dots,i-j+1$, where the
indices of $u_i$ are reduced modulo $n$. Expanding the determinant
of $D$ by minors along the last row we obtain
\begin{equation}
 \label{eq:RecDelta}
\Delta_u(j,i)=\Delta_u(j,i-1)+u_i^2\Delta_u(j,i-2)
\end{equation}
assuming $\Delta_u(j,i)=1$ if $i-j\in\{-2,-1\}$. Comparing the
latter recursive identity with (\ref{eq:dC1}) it is easy to see that
\begin{equation}
 \label{eq:Delta}
\Delta_u(1,i)=C_{i+2}^2(u)\enspace.
\end{equation}
Moreover, from the definition of the determinant it also follows
that
\begin{equation}
 \label{eq:Delta2l}
\Delta_u(1,i)^{2^{tl}}=\Delta_u(1+t,i+t)\quad\mbox{for}\quad 0\leq
t\leq n-1\enspace.
\end{equation}

Now assume $d=\gcd(l,k)=1$ and consider $V$ having the form of
(\ref{eq:V}) as a function of $x\in\mathrm{GF}(2^k)^{**}$ denoted
$V(x)$. It is interesting that $V(x)$ is closely related to
polynomial mappings $q^{(\epsilon)}(x)$ defined in (\ref{eq:q(x)}).
In particular, this connection leads to new properties of
$q^{(\epsilon)}(x)$, with $\epsilon\equiv l'\ (\bmod\;2)$, when it is
not a permutation. Denote
\[\mathcal{T}_i=\{x\in\mathrm{GF}(2^k)^{**}\ |\ {\rm
Tr}_k(x)=i\}\quad\mbox{for}\quad i=0,1\] and let $V(\mathcal{T}_i)$
and $q^{(\epsilon)}(\mathcal{T}_i)$ denote multisets containing all
elements (with repetitions) in the image of $\mathcal{T}_i$ under the
corresponding mapping $V(x)$ or $q^{(\epsilon)}(x)$.

\begin{corollary}
Take $\epsilon\equiv l'\ (\bmod\;2)$. Then
$q^{(\epsilon)}(\mathcal{T}_i)=V(\mathcal{T}_0)$, where $i\equiv k\
(\bmod\;2)$, and $q^{(\epsilon)}(x)$ defines a $3$-to-$1$ mapping on
$\mathcal{T}_i$. Also, if $k$ is odd then
$q^{(0)}(\mathcal{T}_0)=V(\mathcal{T}_1)$ and $q^{(0)}(x)$ defines an
injective mapping on $\mathcal{T}_0$.
\end{corollary}

\begin{pf}
It follows from the proof of Proposition~\ref{pr:ZeC} and
Corollary~\ref{co:ZeZ} that $V(x)$ defines a $3$-to-$1$ mapping on
$\mathcal{T}_0$ and is injective on $\mathcal{T}_1$. This {\em does
not} mean that $V(\mathcal{T}_i)\subset\mathcal{T}_i$ for $i=0,1$ in
the sense of a normal subset relation.

Taking any $x_0\in\mathrm{GF}(2^k)^{**}$ denote
$\Delta=(x_0^{2^l-1}+x_0^{-1})^{-1/(2^l-1)}$ and $\lambda=x_0\Delta$.
It is easy to check that $\lambda+\lambda^{2^l}=\Delta^{2^l}$. Thus,
${\rm Tr}_k(\Delta)=0$ and $\Delta$ defines a $2$-to-$1$ mapping of
$\mathrm{GF}(2^k)^{**}$ on $\mathcal{T}_0$ if $k$ is odd and on
$\mathcal{T}_0\bigcup\{1\}$ if $k$ is even. Therefore,
\begin{align*}
&V(\mathcal{T}_0)=\left\{V\big((x_0^{2^l-1}+x_0^{-1})^{-\frac{1}{2^l-1}}\big)
\ |\ x_0\in\mathrm{GF}(2^k)^{**},\ x_0^{2^l}+x_0\neq 1\right\}\quad\mbox{and}\\
&V\big((x_0^{2^l-1}+x_0^{-1})^{-\frac{1}{2^l-1}}\big)=
\frac{x_0^{2^l}(x_0^{2^l}+1)}{(x_0^{2^l}+x_0+1)^{2^l+1}}=
\frac{\sum_{i=1}^{l'}z_0^{2^{il}}+l'}{z_0^{2^l+1}}=q^{(\epsilon)}(z_0)\enspace,
\end{align*}
where $z_0=x_0^{2^l}+x_0+1\in\mathcal{T}_i$ with $\epsilon\equiv l'\
(\bmod\;2)$ and $i\equiv k\ (\bmod\;2)$. Note that $l'$ and $k$ can
not be both even and $\epsilon=i=0$ is impossible. Here we used that
$V(x)=x^{1-2^l}/(x^{1-2^l}+1)^{2^l+1}$ which is easily obtained from
(\ref{eq:V}). Finally,
\[q^{(\epsilon)}(\mathcal{T}_i)=\left\{q^{(\epsilon)}\big(x_0^{2^l}+x_0+1\big)\ |\
x_0\in\mathrm{GF}(2^k)^{**},\ x_0^{2^l}+x_0\neq 1\right\}\] and
$x^{2^l}+x+1$ defines a $2$-to-$1$ mapping of $\mathrm{GF}(2^k)^{**}$
on $\mathcal{T}_1$ if $k$ is odd and on $\mathcal{T}_0\bigcup\{0\}$
if $k$ is even.

Define $T_l(x)=\sum_{i=0}^{l-1}x^{2^i}$ on $\mathrm{GF}(2^k)$ that is
a permutation polynomial if $l$ is odd and $2$-to-$1$ mapping if $l$
is even. This follows from the fact that $T_l(x)$ is linearized and
$T_l(x)=0$ has the only solution $x=0$ if $l$ is odd and two
solutions $x=0,1$ if $l$ is even (note that
$(T_l(x)+1)T_l(x)=x+x^{2^l}$). Therefore, $T_l(x)$ is a permutation
of $\mathcal{T}_0$ for any $l$ and odd $k$ since ${\rm
Tr}_k(T_l(x))=0$ if $l$ is even (in this case $T_l(x)=T_l(x+1)$) and
is equal to ${\rm Tr}_k(x)$ if $l$ is odd.

Dickson polynomial number $2^l+1$ can be written as
$D_{2^l+1}(x)=x^{2^l+1}(1+T_l(x^{-1})^2)$ (see, for instance,
\cite{DiDo04}). If $k$ is odd then $\gcd(2^l+1,2^k-1)=1$ and, by
\cite{DiDo04}, $D_{2^l+1}(x)$ is a permutation on $\mathcal{H}_0$.
Thus, $D_{2^l+1}(x^{-1})^{-1}$ is a permutation on $\mathcal{T}_0$.
Note that
\[D_{2^l+1}(x^{-1})^{-1}=\frac{x^{2^l+1}}{T_l(x)^2+1}=
\frac{x}{T_l(x)+1}+\left(\frac{x}{T_l(x)+1}\right)^2+x\enspace.\]
Therefore, if $k$ is odd then
\begin{align*}
&V(\mathcal{T}_1)=\bigg\{\frac{(T_l(x)+1)^{2^{2k}+1}}{\left(T_l(x)+T_l^{2^k}(x)\right)^{2^k+1}}
\ |\ x_0\in\mathcal{T}_0\bigg\}\quad\mbox{and}\\
&q^{(0)}(T_l(x_0))=
\end{align*}
\end{pf}

\section{Zeros of $P_a(x)$ when $\gcd(l,k)=1$}
 \label{sec:zer1}
In this section, we analyze the zeros in $\mathrm{GF}(2^k)$ of the
polynomial $P_a(x)$ assuming that $l$ and $k$ are coprime integers
with $l<k$. In this case, denote $l'=l^{-1}\ (\bmod\;k)$ and take
$R(x)$ defined in (\ref{eq:R}). The following Lemma~\ref{le:V}
easily follows from the earlier mentioned fundamental result on
permutation polynomials due to Dobbertin.

Also note the fact that since $l'l\equiv 1\ (\bmod\;k)$ then
\[(2^l-1)(1+2^l+2^{2l}+\cdots+2^{(l'-1)l})=2^{ll'}-1\equiv 1\
(\bmod\;2^k-1)\enspace.\] Therefore, $u^{2^{l'l}}=u^2$ for any
$u\in\mathrm{GF}(2^k)$ and this identity will be used repeatedly
further in the proofs.

\begin{lemma}
 \label{le:V}
Take $F_a(x)$ defined in (\ref{eq:F}). Then for any
$a\in\mathrm{GF}(2^k)^*$, the element $\mathcal{V}=R(a^{-1})$ is a
zero of $F_a(x)$ in $\mathrm{GF}(2^k)$.
\end{lemma}

\begin{pf}
Since $q(x)$ from (\ref{eq:q(x)}) is a permutation polynomial on
$\mathrm{GF}(2^k)^*$, then for any fixed $a\in\mathrm{GF}(2^k)^*$ the
equation
\begin{equation}
 \label{eq:Dob}
ax^{2^l+1}=\sum_{i=1}^{l'}x^{2^{il}}+l'+1
\end{equation}
has exactly one solution $\mathcal{V}=R(a^{-1})$ in
$\mathrm{GF}(2^k)^*$. Raising (\ref{eq:Dob}) to the power of $2^l$
results in
\[a^{2^l}x^{2^{2l}+2^l}=\sum_{i=2}^{l'+1}x^{2^{il}}+l'+1=\sum_{i=2}^{l'}x^{2^{il}}+x^{2^{l+1}}+l'+1\enspace.\]
The latter identity, after being added to (\ref{eq:Dob}) and setting
$x=\mathcal{V}$, gives
\[a\mathcal{V}^{2^l+1}=a^{2^l}\mathcal{V}^{2^{2l}+2^l}+\mathcal{V}^{2^l}+\mathcal{V}^{2^{l+1}}\]
and consecutively, since $\mathcal{V}\neq 0$,
$F_a(\mathcal{V})=a^{2^l}\mathcal{V}^{2^{2l}}+\mathcal{V}^{2^l}+a\mathcal{V}+1=0$.\qed
\end{pf}

Now we introduce a particular sequence of polynomials over
$\mathrm{GF}(2^k)$ and prove some important properties of these that
will be used further for getting the main result of this section
about zeros of $P_a(x)$. Denote
\[e(i)=1+2^l+2^{2l}+\cdots+2^{(i-1)l},\quad\mbox{for}\quad i=1,\dots,l'\]
so, in particular, $e(l')=(2^l-1)^{-1}\ (\bmod\;2^k-1)$. Now take
every additive term $x^e$ with $e\neq 0$ in the polynomial
$1+(1+x)^{e(i)}$ and replace the exponent $e$ with the cyclotomic
equivalent number obtained by shifting the binary expansion of $e$
maximally (till you get an odd number) in the direction of the least
significant bits. We call this {\em reduction} procedure. Recall
that two exponents $e_1$ and $e_2$ are cyclotomic equivalent if $2^i
e_1\equiv e_2\ (\bmod\;2^k-1)$ for some $i<k$. For instance,
$x^{2^{il}}$ is reduced to $x$ and $x^{2^{il}+2^{jl}}$ is reduced to
$x^{1+2^{(j-i)l}}$ if $i<j$ and so on. The obtained reduced
polynomials are denoted as $H_i(x)$ and we use square brackets to
denote application of the described reduction procedure to a
polynomial, so $H_i(x)=[1+(1+x)^{e(i)}]$ for $i=1,\dots,l'$. The
first few polynomials in the sequence (after eliminating all pairs
of equal terms) are
\begin{eqnarray*}
H_1(x)&=&x\\
H_2(x)&=&[x+x^{2^l}+x^{1+2^l}]=x+x+x^{1+2^l}=x^{1+2^l}\\
H_3(x)&=&[x+x^{2^l}+x^{2^{2l}}+x^{1+2^l}+x^{1+2^{2l}}+x^{2^l+2^{2l}}+x^{1+2^l+2^{2l}}]\\
&=&x+x+x+x^{1+2^l}+x^{1+2^{2l}}+x^{1+2^l}+x^{1+2^l+2^{2l}}\\
&=&x+x^{1+2^{2l}}+x^{1+2^l+2^{2l}}\enspace.
\end{eqnarray*}

\begin{lemma}
 \label{le:H}
If polynomials $H_i(x)$ are defined as above then
\[{\rm Tr}_k(H_i(x))={\rm Tr}_k\big(1+(1+x)^{e(i)}\big)\]
for any $x\in\mathrm{GF}(2^k)$ and $i=1,\dots,l'$. Also let
\[Q(x)=(x_0^{2^l+1}+x_0)x^{2^l}+x_0^2 x+x_0\] for any
$x_0\in\mathrm{GF}(2^k)^*$. Then
\[Q(H_{l'}(x_0^{-1}))=(1+x_0)(1+x_0^{-1})^{e(l')}\enspace.\]
\end{lemma}

\begin{pf}
Obviously, we get the trace identity for $H_{l'}(x)$ from the
definition. Further,
\begin{eqnarray*}
H_i(x)&=&[1+(1+x)^{e(i)}]\\
&=&[1+(1+x)^{e(i-1)}(1+x)^{2^{(i-1)l}}]\\
&=&[H_{i-1}(x)+x^{2^{(i-1)l}}(1+x)^{e(i-1)}]\\
&\stackrel{(\ast)}{=}&x(1+x)^{e(i)-1}+H_{i-1}(x)\enspace,
\end{eqnarray*}
where $(\ast)$ follows from the following argumentation. First, note
that the exponents of additive terms in $x(1+x)^{e(i)-1}$ are
exactly all $2^{i-1}$ distinct integers of the form $1+t_1
2^l+\cdots+t_{i-1} 2^{(i-1)l}$ with $t_j\in\{0,1\}$ for
$j=1,\dots,i-1$ and the reduction does not apply to any of these so
\[[x(1+x)^{e(i)-1}]=x(1+x)^{e(i)-1}\enspace.\]
On the other hand, the number of terms in
$[x^{2^{(i-1)l}}(1+x)^{e(i-1)}]$ is also equal to $2^{i-1}$ since
the exponents in these terms are exactly all the integers of the
form $t_0+t_1 2^l+\cdots+t_{i-2} 2^{(i-2)l}+2^{(i-1)l}$ with
$t_j\in\{0,1\}$ for $j=0,\dots,i-2$ and none of these become equal
after the reduction. Moreover, every such an exponent, after
reduction, can be found in $x(1+x)^{e(i)-1}$ so
\[[x^{2^{(i-1)l}}(1+x)^{e(i-1)}]=x(1+x)^{e(i)-1}\enspace.\]
Also note that all terms of $H_{i-1}(x)$ are also present in
$x(1+x)^{e(i)-1}$. Thus, the number of terms in $H_i(x)$ that remain
after eliminating all pairs of equal terms and denoted as $\#H_i$ is
equal to $2^{i-1}-\#H_{i-1}$. Unfolding the obtained recursive
expression for $H_i(x)$ starting from $H_1(x)=x$ we get that
\begin{equation}
 \label{eq:H}
H_i(x)=x(1+(1+x)^{2^l}+(1+x)^{2^l+2^{2l}}+\cdots+(1+x)^{e(i)-1})\enspace.
\end{equation}

Now we can evaluate
\begin{align*}
&Q(H_{l'}(x_0^{-1}))\\
&=(x_0^{2^l+1}+x_0)H_{l'}(x_0^{-1})^{2^l}+x_0^2 H_{l'}(x_0^{-1})+x_0\\
&=(x_0+x_0^{-2^l+1})
\left(1+(1+x_0^{-1})^{2^{2l}}+(1+x_0^{-1})^{2^{2l}+2^{3l}}+\cdots+(1+x_0^{-1})^{2^{2l}+\cdots+2^{l'l}}\right)\\
&\quad{}+x_0\left(1+(1+x_0^{-1})^{2^l}+(1+x_0^{-1})^{2^l+2^{2l}}+\cdots+(1+x_0^{-1})^{e(l')-1}\right)+x_0\\
&=\left((x_0+x_0^{-2^l+1})+x_0(1+x_0^{-1})^{2^l}\right)
\left(1+(1+x_0^{-1})^{2^{2l}}+\cdots+(1+x_0^{-1})^{2^{2l}+\cdots+2^{(l'-1)l}}\right)\\
&\quad{}+(x_0+x_0^{-2^l+1})(1+x_0^{-1})^{2^{2l}+\cdots+2^{l'l}}+x_0+x_0\\
&=x_0(1+x_0^{-1})^{2^l+2^{2l}+\cdots+2^{l'l}}\\
&=x_0(1+x_0^{-1})^{2+2^l+2^{2l}+\cdots+2^{(l'-1)l}}\\
&=(1+x_0)(1+x_0^{-1})^{e(l')}
\end{align*}
as claimed.\qed
\end{pf}

\begin{lemma}
 \label{le:Trv0}
For any $a\in\mathrm{GF}(2^k)^*$ let $x_0\in\mathrm{GF}(2^k)$
satisfy $x_0^{2^l+1}+x_0=a$. Then
\[{\rm Tr}_k\big(1+(1+x_0^{-1})^{e(l')}\big)={\rm Tr}_k(R(a^{-1}))\enspace.\]
\end{lemma}

\begin{pf}
Denote $\Gamma=x_0^{2^l-1}+x_0^{-1}$ (obviously $\Gamma\neq 0$ since
$x_0\neq 1$), $\Delta=\Gamma^{-e(l')}$, and further, using
Lemma~\ref{le:H}, evaluate
\[Q(H_{l'}(x_0^{-1}))x_0^{e(l')}=(1+x_0)(1+x_0)^{e(l')}=(1+x_0^{2^l})^{e(l')}\]
and thus, $Q(H_{l'}(x_0^{-1}))^{2^l-1}=\Gamma$ or, equivalently,
\begin{equation}
 \label{eq:z0}
Q(H_{l'}(x_0^{-1}))=\Delta^{-1}\enspace.
\end{equation}

In what follows, we use the technique suggested by Dobbertin for
proving \cite[Theorem~1]{Do99}. Take the polynomial $F_a(x)$ defined
in (\ref{eq:F}) and note that
\begin{align*}
F_a(x)&=a^{2^l}x^{2^{2l}}+x^{2^l}+ax+1\\
&=a^{2^l}x^{2^{2l}}+x_0^{2^{l+1}}x^{2^l}+x_0^{2^l}+(x_0^{2^l-1}+x_0^{-1})\left((x_0^{2^l+1}+x_0)x^{2^l}+x_0^2 x+x_0\right)\\
&=Q(x)^{2^l}+\Gamma Q(x)=Q(x)\left(Q(x)^{2^l-1}+\Delta^{-(2^l-1)}\right)
\end{align*}
for $x_0^{2^l+1}+x_0=a$ and, therefore, by (\ref{eq:z0}),
$F_a(H_{l'}(x_0^{-1}))=0$. Consider the equation
\begin{equation}
 \label{eq:QDel}
Q(x)+\Delta^{-1}=0
\end{equation}
whose roots are also the zeros of $F_a(x)$. We will show that
(\ref{eq:QDel}) has exactly two roots with $H_{l'}(x_0^{-1})$ and
$R(a^{-1})$ being among them (however, we do not claim that
$R(a^{-1})\neq H_{l'}(x_0^{-1})$). Multiplying (\ref{eq:QDel}) by
$\mu=(x_0^2\Delta)^{-1}$ and using that
$(x_0^{2^l+1}+x_0)\Delta^{2^l-1}=x_0^2$ gives
\[\mu((x_0^{2^l+1}+x_0)x^{2^l}+x_0^2 x+x_0+\Delta^{-1})=
(x/\Delta)^{2^l}+x/\Delta+x_0\mu+x_0^2\mu^2=0\enspace,\] which has
exactly two solutions $z_0=H_{l'}(x_0^{-1})$ (see (\ref{eq:z0})) and
$z_1=H_{l'}(x_0^{-1})+\Delta$, since its linearized homogeneous part
$(x/\Delta)^{2^l}+x/\Delta$ has exactly two roots $x=0$ and
$x=\Delta$. Thus
\[z_0+z_1=\Delta=\left(\frac{x_0}{1+x_0^{2^l}}\right)^{e(l')}\enspace.\]
Using $(x_0^{2^l}+1)\Delta^{2^l-1}=x_0$ it is easy to see that
$\Delta^{2^l}=x_0\Delta+(x_0\Delta)^{2^l}$ and we have ${\rm
Tr}_k(\Delta)=0$.

Now we show that none of the possible roots of $Q(x)=0$ is a
solution of (\ref{eq:Dob}). In fact, suppose that $Q(z)=0$. Then,
since $x_0\neq 0$, we have $z^{2^l}=(x_0 z)^{2^l}+x_0 z+1$ and
$az^{2^l}=x_0^2 z+x_0$ (since $a=x_0^{2^l+1}+x_0$). We put such a
$z$ into (\ref{eq:Dob}) and compute
\begin{eqnarray*}
&&az^{2^l+1}+\sum_{i=1}^{l'}z^{2^{il}}+l'+1\\
&&=(x_0^2 z+x_0)z+\sum_{i=0}^{l'-1}(x_0 z)^{2^{il}}+\sum_{i=1}^{l'}(x_0 z)^{2^{il}}+l'+l'+1\\
&&=1\enspace.
\end{eqnarray*}

Therefore, recalling the proved identity
$F_a(x)=Q(x)(Q(x)^{2^l-1}+\Delta^{-(2^l-1)})$ and keeping in mind
that $\gcd(2^l-1,2^k-1)=1$ we see that $\mathcal{V}=R(a^{-1})$ which
is the unique solution of (\ref{eq:Dob}) and, by Lemma~\ref{le:V},
also the root of $F_a(x)=0$, satisfies $Q(\mathcal{V})=\Delta^{-1}$.
Recall that (\ref{eq:QDel}) has exactly two solutions
$z_0=H_{l'}(x_0^{-1})$ and $z_1=H_{l'}(x_0^{-1})+\Delta$. Thus,
$R(a^{-1})+H_{l'}(x_0^{-1})=\Delta$ or $R(a^{-1})=H_{l'}(x_0^{-1})$
(although we do not need in our proof that $R(a^{-1})\neq
H_{l'}(x_0^{-1})$, we believe that this holds) and, by
Lemma~\ref{le:H}
\[{\rm Tr}_k(R(a^{-1}))={\rm Tr}_k(H_{l'}(x_0^{-1}))={\rm Tr}_k(1+(1+x_0^{-1})^{e(l')})\]
as claimed.\qed
\end{pf}

\begin{theorem}
 \label{th:Pa}
For any $a\in\mathrm{GF}(2^k)^*$ and a positive integer $l<k$ with
$\gcd(l,k)=1$ polynomial $P_a(x)$ has either none, one, or three
zeros in $\mathrm{GF}(2^k)$. Further, $P_a(x)$ has exactly one zero
in $\mathrm{GF}(2^k)$ if and only if ${\rm Tr}_k(R(a^{-1})+1)=1$,
where $R(x)$ is defined in (\ref{eq:R}). Moreover, if $P_a(x_0)=0$
for some $x_0\in\mathrm{GF}(2^k)$ then
\[{\rm Tr}_k(R(a^{-1}))={\rm Tr}_k(H_{l'}(x_0^{-1}))={\rm Tr}_k((1+x_0^{-1})^{e(l')}+1)\]
where polynomials $H_i(x)$ are defined in (\ref{eq:H}). Finally, the
following distribution holds for $k$ odd (resp. $k$ even)
\[\begin{array}{llll}
M_0&=&\frac{2^k+1}{3}&\quad (\mbox{resp.}\ \frac{2^k-1}{3})\\
M_1&=&2^{k-1}-1&\quad (\mbox{resp.}\ 2^{k-1})\\
M_3&=&\frac{2^{k-1}-1}{3}&\quad (\mbox{resp.}\
\frac{2^{k-1}-2}{3})\enspace.
\end{array}\]
\end{theorem}

\begin{pf}
Assume $P_a(x_0)=0$ for some $x_0\in\mathrm{GF}(2^k)$. Now we
substitute $x$ in $P_a(x)$ with $x+x_0$ to get
\[(x+x_0)^{2^l+1}+(x+x_0)+a=0\]
or
\[x^{2^l+1}+x_0 x^{2^l}+x_0^{2^l}x+x_0^{2^l+1}+x+x_0+a=0\]
which implies
\[x^{2^l+1}+x_0 x^{2^l}+(x_0^{2^l}+1)x=0\enspace.\]
Since $x=0$ corresponds to $x_0$ being the zero of $P_a(x)$, we can
divide the latter equation by $x$. Further, after substituting
$y=x^{-1}$ we note that $P_a(x)$ has $i$ zeros if and only if the
reciprocal equation, given by
\begin{equation}
 \label{eq:yeq}
(x_0^{2^l}+1)y^{2^l}+x_0 y+1=0
\end{equation}
has $i-1$ zeros. This affine equation has either zero roots in
$\mathrm{GF}(2^k)$ or the same number of roots as its homogeneous
part $(x_0^{2^l}+1)y^{2^l}+x_0 y$ which is seen to have exactly two
solutions, the zero solution and a unique nonzero solution, since
$\gcd(2^l-1,2^k-1)=1$. Therefore, it can be concluded that
$P_a(x)=0$ can have either zero, one, or three solutions in
$\mathrm{GF}(2^k)$.

Now we need to find the conditions when there exists a solution of
(\ref{eq:yeq}). Let $y=tw$, where $t^{2^l-1}=c$ and
$c=\frac{x_0}{x_0^{2^l}+1}$. Since $\gcd(2^l-1,2^k-1)=1$, there is a
one-to-one correspondence between $t$ and $c$. Then (\ref{eq:yeq})
is equivalent to
\[w^{2^l}+w+\frac{1}{ct(x_0^{2^l}+1)}=0\enspace.\]
Hence, (\ref{eq:yeq}) has no solutions if and only if
\[{\rm Tr}_k\left(\frac{1}{ct(x_0^{2^l}+1)}\right)=1\enspace.\]
This easily follows from the fact that the linear operator
$L(\omega)=\omega^{2^l}+\omega$ on $\mathrm{GF}(2^k)$ has the kernel
of dimension one and, thus, the number of elements in the image of
$L$ is $2^{k-1}$. For any $\omega\in\mathrm{GF}(2^k)$, we have ${\rm
Tr}_k(\omega^{2^l}+\omega)=0$ leading to the conclusion that the
image of $L$ contains all the elements in $\mathrm{GF}(2^k)$ having
trace zero since the total number of such elements in
$\mathrm{GF}(2^k)$ is exactly $2^{k-1}$.

Since $c=t^{2^l-1}$ then $t=c^{1+2^l+2^{2l}+\cdots+2^{(l'-1)l}}$.
Thus, from the definition of $c$ and $t$ we get
\begin{eqnarray*}
{\rm Tr}_k\left(\frac{1}{ct(x_0^{2^l}+1)}\right)
&=&{\rm Tr}_k\left(\bigg(\frac{x_0^{2^l}+1}{x_0}\bigg)^{1+e(l')}\left(\frac{1}{x_0^{2^l}+1}\right)\right)\\
&=&{\rm Tr}_k\Bigg(\frac{(x_0^{2^l}+1)^{e(l')}}{x_0^{1+e(l')}}\Bigg)\\
&=&{\rm Tr}_k\Bigg(\frac{(x_0+1)^{2^le(l')}}{x_0^{2^le(l')}}\Bigg)\\
&=&{\rm Tr}_k\big((1+x_0^{-1})^{e(l')}\big)\enspace.
\end{eqnarray*}

We conclude that $P_a(x)$ has exactly one zero in $\mathrm{GF}(2^k)$
(which is $x_0$) if and only if
\begin{equation}
 \label{eq:M2}
{\rm Tr}_k\big((1+x_0^{-1})^{e(l')}\big)=1
\end{equation}
or, equivalently, for all such $a$ that $a=x_0^{2^l+1}+x_0$ with
(\ref{eq:M2}) holding. Combining this with the result of
Lemma~\ref{le:Trv0}, we conclude that $P_a(x)$ has exactly one zero
in $\mathrm{GF}(2^k)$ if and only if
\[{\rm Tr}_k(R(a^{-1})+1)=1\]
(the ``if" part follows from the fact that $R(x)$ is a permutation
polynomial and seeing the value of $M_1$ that is computed in the
next paragraph). In the case of none or three zeros, ${\rm
Tr}_k(R(a^{-1})+1)=0$. The trace identities follow from
Lemmas~\ref{le:H} and \ref{le:Trv0}.

Now note that since $e(l')=1+2^l+2^{2l}+\cdots+2^{(l'-1)l}$ is
invertible modulo $2^k-1$ with the multiplicative inverse equal to
$2^l-1$ then $\gcd(e(l'),2^k-1)=1$ and thus,
$x\mapsto(1+x^{-1})^{e(l')}$ is a one-to-one mapping of
$\mathrm{GF}(2^k)^*$ onto $\mathrm{GF}(2^k)\setminus\{1\}$.
Therefore, if $k$ is odd (resp. $k$ is even) then the number of
$x_0\in\mathrm{GF}(2^k)^*$ satisfying (\ref{eq:M2}) is equal to
$2^{k-1}-1$ (resp. $2^{k-1}$) and obviously $x_0\neq 1$. This also
gives the value of $M_1$ since every $x_0$ satisfying (\ref{eq:M2})
provides a unique $a=x_0^{2^l+1}+x_0\in\mathrm{GF}(2^k)^*$ such that
$P_a(x)$ has exactly one zero. Now note that if $a=0$ then
$P_a(x)=x^{2^l+1}+x+a$ has exactly two zeros $x=\{0,1\}$. Thus,
considering the mapping $x\mapsto x^{2^l+1}+x$ for $x$ running
through $\mathrm{GF}(2^k)\setminus\{0,1\}$ it is easy to see that
$M_1+3M_3=2^k-2$ and, knowing $M_1$ we can find $M_3$. Finally, the
last remaining unknown $M_0$ can be evaluated from the obvious
equation $M_0+M_1+M_3=|\mathrm{GF}(2^k)^*|=2^k-1$.\qed
\end{pf}

Note that Bluher in \cite[Theorem~5.6]{Bl04} (see
Theorem~\ref{th:Bl} below), in particular, found the possible number
of zeros of $P_a(x)$ and calculated the corresponding values of
$M_i$, in the notations of our Theorem~\ref{th:Pa}. This was also
done earlier for odd $k$ in \cite[Lemma~9]{HeZi01}.

\section{Zeros of $P_a(x)$ when $\gcd(l,k)\geq 1$}
 \label{sec:zer2}
In this section, we analyze the zeros in $\mathrm{GF}(2^k)$ of the
polynomial $P_a(x)$ assuming that $l,k$ are positive integers with
$l<k$ and $\gcd(l,k)=d\geq 1$. In this case, let $k=nd$ for some
$n>1$ and also recall our notation $u_i=u^{2^{il}}$ for any
$u\in\mathrm{GF}(2^k)$ and $i=0,\dots,n-1$. First, keep in mind the
following result that can be obtained combining Theorems~5.6 and 6.4
in \cite{Bl04}.

\begin{theorem}[\cite{Bl04}]
 \label{th:Bl}
For any $b\in\mathrm{GF}(2^k)^*$, take polynomials
\[f(x)=x^{2^l+1}+b^2 x+b^2\quad{\mbox and}\quad
g(x)=b^{-1}f(bx^{2^l-1})=b^{2^l}x^{2^{2l}-1}+b^2 x^{2^l-1}+b\] over
$\mathrm{GF}(2^k)$ and let $\gcd(l,k)=d$. Then exactly one of the
following holds
\renewcommand{\theenumi}{\roman{enumi}}
\renewcommand{\labelenumi}{(\theenumi)}
\begin{enumerate}
\item\label{it:1} $f(x)$ has none or two zeros in
    $\mathrm{GF}(2^k)$ and $g(x)$ has none zeros in
    $\mathrm{GF}(2^k)$;

\item\label{it:2} $f(x)$ has one zero in $\mathrm{GF}(2^k)$,
    $g(x)$ has $2^d-1$ zeros in $\mathrm{GF}(2^k)$ and each
    rational root $\delta$ of $g(x)$ satisfies ${\rm
    Tr}^k_d\left(b^{-1}\delta^{-(2^l+1)}\right)\neq 0$;

\item\label{it:3} $f(x)$ has $2^d+1$ zeros in
    $\mathrm{GF}(2^k)$, $g(x)$ has $2^{2d}-1$ zeros in
    $\mathrm{GF}(2^k)$ and each rational root $\delta$ of $g(x)$
    satisfies ${\rm
    Tr}^k_d\left(b^{-1}\delta^{-(2^l+1)}\right)=0$.
\end{enumerate}
Let $T_i$ denote the number of $b\in\mathrm{GF}(2^k)^*$ such that
$f(x)=0$ has exactly $i$ roots in $\mathrm{GF}(2^k)$. Then the
following distribution holds for $k/d$ odd (resp. $k/d$ even)
\[\begin{array}{llll}
T_0&=&\frac{(2^k+1)2^{d-1}}{2^d+1}&\quad ({\mbox resp.}\
\frac{(2^k-1)2^{d-1}}{2^d+1})\ ,\\
T_1&=&2^{k-d}-1&\quad (\mbox{resp.}\ 2^{k-d})\ ,\\
T_2&=&\frac{(2^k-1)(2^{d-1}-1)}{2^d-1}&\quad (\mbox{in both cases})\ ,\\
T_{2^d+1}&=&\frac{2^{k-d}-1}{2^{2d}-1}&\quad (\mbox{resp.}\
\frac{2^{k-d}-2^d}{2^{2d}-1})\enspace.
\end{array}\]
\end{theorem}

\begin{nnote}
 \label{no:1}
Take a linearized polynomial
\begin{equation}
 \label{eq:L}
L_a(x)=a^{2^l}x^{2^{2l}}+x^{2^l}+ax\enspace.
\end{equation}
Note that zeros in $\mathrm{GF}(2^k)$ of $L_a(x)$ form a vector
subspace over $\mathrm{GF}(2^d)$ and thus, the number of zeros can
be equal to $1,2^d,2^{2d},\dots,2^{2l}$ (we will see that, in fact,
$L_a(x)$ can not have more than $2^{2d}$ zeros). Assume $a\neq 0$,
then dividing $L_a(x)$ by $a_0 a_1 x$ (we remove one zero $x=0$) and
then substituting $x$ with $a_0^{-1}x$ leads to $a_1^{-2^l}
x^{2^{2l}-1}+a_1^{-2}x^{2^l-1}+a_1^{-1}$ which has the form of
polynomial $g(x)$ from Theorem~\ref{th:Bl} taking $b=a_1^{-1}$ (note
a $1$-to-$1$ correspondence between $a$ and $b$). This leads to the
corresponding $f(x)=x^{2^l+1}+a_1^{-2}x+a_1^{-2}$. Finally,
substituting $x$ in the latter $f(x)$ with $a_0^{-2}x$ and
multiplying by $a_0^2 a_1^2$ we get $x^{2^l+1}+x+a_0^2=P_{a^2}(x)$.
By Theorem~\ref{th:Bl}, we obtain the relation between the number of
zeros of $L_a(x)$ and $P_{a^2}(x)$. We also conclude that $P_a(x)$
has either $0$, $1$, $2$ or $2^d+1$ zeros in $\mathrm{GF}(2^k)$ and
$M_i=T_i$ for $i=0,1,2,2^d+1$. It can be checked directly that
$L_a(x)=0$ for some $x\neq 0$ if and only if $P_{a^2}(a_0
x^{2^l-1})=0$.
\end{nnote}

\begin{proposition}
 \label{pr:P02Zero}
Take any $a\in\mathrm{GF}(2^k)^*$. Then polynomial $P_a(x)$ has none
or exactly two zeros in $\mathrm{GF}(2^k)$ if and only if
$Z_n(a)\neq 0$. Also if $n$ is odd (resp. $n$ is even) then
\[\begin{array}{llll}
M_0&=&\frac{(2^k+1)2^{d-1}}{2^d+1}&\quad (\mbox{resp.}\
\frac{(2^k-1)2^{d-1}}{2^d+1})\ ,\\
M_2&=&\frac{(2^k-1)(2^{d-1}-1)}{2^d-1}&\quad (\mbox{in both
cases})\enspace.
\end{array}\]
\end{proposition}

\begin{pf}
Consider the equation $L_a(x)=a_1 x_2+x_1+a_0 x_0=0$ and show that
in our case, it has the only zero solution. Taking $L_a(x)=0$ and
all its $2^{il}$ powers we obtain $n$ equations
\[L_a^{2^{il}}(x)=a_{i+1}x_{i+2}+x_{i+1}+a_i x_i=0\quad\mbox{for}\quad
i=0,\dots,n-1\enspace,\] where all indices are calculated modulo
$n$. If $x_i$ $(i=0,\dots,n-1)$ are considered as independent
variables then the obtained system of $n$ linear equations with $n$
unknowns has the following matrix with the antidiagonal structure,
assuming $n>2$
\begin{equation}
 \label{eq:matr}
\left(\begin{array}{cccccc}
0&0&\cdots&a_1&1&a_0\\
0&&\adots&1&a_1&0\\
\vdots&\adots&\adots&\adots&&\vdots\\
a_{n-2}&1&\adots&\adots&&0\\
1&a_{n-2}&&&0&a_{n-1}\\
a_{n-1}&0&\cdots&0&a_0&1
\end{array}\right)\enspace.
\end{equation}
If $n=2$ then $l=d$ and $L_a(x)=x_1+(a_0 + a_1)x_0$. The
corresponding matrix is $\mathcal{M}_2=\left(\begin{array}{cc}
1&a_0 + a_1\\
a_0 + a_1&1
\end{array}\right)$ having the determinant equal to
\[1+a_0^2 + a_1^2=Z_2^2(a)\neq 0\enspace.\]
Let the columns of (\ref{eq:matr}) be numbered from $1$ to $n>2$.
Permuting the columns in (\ref{eq:matr}) (reorder them as
$n-1,n-2,\dots,1,n$) we obtain a symmetric three-diagonal cyclic
matrix $\mathcal{M}_n$ containing ones on the main diagonal, with
\[\mathcal{M}_n(i,i+1)=\mathcal{M}_n(i+1,i)=a_i\quad\mbox{for}\quad
i=1,\dots,n-1\] and corner elements
$\mathcal{M}_n(1,n)=\mathcal{M}_n(n,1)=a_0$. If
$\vec{x}=(x_1,\dots,x_{n-1},x_0)^{\rm T}$ and
$\vec{0}=(0,\dots,0)^{\rm T}$ then the system has the following
matrix representation
\begin{equation}
 \label{eq:syst}
\mathcal{M}_n\vec{x}=\vec{0}\enspace.
\end{equation}
The determinant of (\ref{eq:matr}) is equal to the determinant of
$\mathcal{M}_n$ and can be computed expanding the latter by minors
along the last row. Doing this it is easy to see that
\begin{eqnarray*}
\det\mathcal{M}_n&=&\Delta_a(1,n-2)+a_{n-1}(a_{n-1}\Delta_a(1,n-3)+a_0\dots a_{n-2})\\
&&\quad\quad\quad\quad\quad\ {}+a_0(a_0\Delta_a(2,n-2)+a_1\dots a_{n-1})\\
&\stackrel{(\ref{eq:Delta},\ref{eq:Delta2l})}{=}&C_n^2(a)+a_{n-1}^2 C_{n-1}^2(a)+(a_0 C_{n-1}^{2^l}(a))^2\\
&\stackrel{(\ref{eq:dC1})}{=}&C_{n+1}^2(a)+(a_0
C_{n-1}^{2^l}(a))^2\\
&\stackrel{(\ref{eq:Z})}{=}&Z_n^2(a)\neq 0\enspace.
\end{eqnarray*}
Thus, (\ref{eq:syst}) has exactly one solution which is
$\vec{x}=\vec{0}$. Now note that every $x\in\mathrm{GF}(2^k)$ with
$L_a(x)=0$ provides a solution to the system given by
$x_i=x^{2^{il}}$ for $i=0,\dots,n-1$. Therefore, if $Z_n(a)\neq 0$
then $L_a(x)=0$ has exactly one root (which is equal to zero). By
Note~\ref{no:1} and Theorem~\ref{th:Bl}~(\ref{it:1}), $P_{a^2}(x)$
has either none or exactly two zeros in $\mathrm{GF}(2^k)$ and the
identities for $M_0$ and $M_2$ follow as well. Finally, note that
$Z_n(a^2)=Z_n^2(a)$ and, therefore, the conditions of the theorem
are satisfied by any $a^{2^i}$ with $i=0,\dots,k-1$.

Using Corollary~\ref{co:ZeZ}, we can obtain the number of
$a\in\mathrm{GF}(2^k)^*$ such that $Z_n(a)\neq 0$ (note that
$Z_n(0)=1$). Observe that this number is identical to $T_0+T_2$
taken from Theorem~\ref{th:Bl} that is equal to the number of
$a\in\mathrm{GF}(2^k)^*$ such that $P_a(x)=0$ has none or exactly
two roots in $\mathrm{GF}(2^k)$ (see Note~\ref{no:1}). Therefore, if
$P_a(x)$ has none or exactly two roots in $\mathrm{GF}(2^k)$ then
$a$ is necessarily such that $Z_n(a)\neq 0$.\qed
\end{pf}

The following proposition provides a criterion to distinguish
between the cases when $P_a(x)$ has none and when it has exactly two
zeros in $\mathrm{GF}(2^k)$.

\begin{proposition}
 \label{pr:P02ZeroD}
Take any $a\in\mathrm{GF}(2^k)^*$. Then polynomial $P_a(x)$ has
exactly two zeros in $\mathrm{GF}(2^k)$ if and only if $Z_n(a)\neq
0$ and ${\rm Tr}_d\left({\rm N}_d^k(a)/Z_n^2(a)\right)=0$. Moreover,
if $d$ is odd then these two zeros are $(W+\mu)Z_n(a)/C_n(a)$ for
$\mu\in\{0,1\}$, where
\[W=\frac{C_{n+1}(a)}{Z_n(a)}+\sum_{i=0}^{\frac{d-1}{2}}
\left(\frac{{\rm N}_d^k(a)}{Z_n^2(a)}\right)^{2^{2i}}\enspace.\]
\end{proposition}

\begin{pf}
First, consider equation $a^{2^l}x^{2^l+1}+x+a=0$. Using the
substitution $x=a^{-1}y$ and multiplying by $a$, the latter equation
is transformed into $P_{a^2}(y)=y^{2^l+1}+y+a^2=0$ having the same
number of roots. Thus, by Proposition~\ref{pr:P02Zero},
$a^{2^l}x^{2^l+1}+x+a$ has none or exactly two zeros if and only if
$Z_n(a^2)=Z_n^2(a)\neq 0$.

We prove by induction that for any $u\in\mathrm{GF}(2^k)$ being a
root of this equation and $i=1,\dots,n$ holds
\[u_i=\frac{u C_{i+1}^2(a)+a(C_i^2(a))^{2^l}}
{a_i\left(u C_i^2(a)+a(C_{i-1}^2(a))^{2^l}\right)}\] assuming
$C_0(x)=0$. For $i=1$ the identity is obvious since $a_1
u^{2^l+1}+u+a=0$. Assuming the identity holds for $i<t$ we get for
$i=t>1$
\begin{eqnarray*}
u_t=u_{t-1}^{2^l}&=&\frac{u_1(C_t^2(a))^{2^l}+a_1(C_{t-1}^2(a))^{2^{2l}}}
{a_t\left(u_1(C_{t-1}^2(a))^{2^l}+a_1(C_{t-2}^2(a))^{2^{2l}}\right)}\\
&=&\frac{(u+a)(C_t^2(a))^{2^l}+u a_1^2(C_{t-1}^2(a))^{2^{2l}}}
{a_t\left((u+a)(C_{t-1}^2(a))^{2^l}+u a_1^2(C_{t-2}^2(a))^{2^{2l}}\right)}\\
&\stackrel{(\ref{eq:dC2})}{=}&\frac{u C_{t+1}^2(a)+a(C_t^2(a))^{2^l}}
{a_t\left(u C_t^2(a)+a(C_{i-1}^2(a))^{2^l}\right)}
\end{eqnarray*}
using induction hypothesis and since $u_1=(u+a)/a_1 u$.

In particular, for $i=n$ we get
\begin{align*}
&u_n=u=\frac{u C_{n+1}^2(a)+a(C_n^2(a))^{2^l}}
{a\left(u C_n^2(a)+a(C_{n-1}^2(a))^{2^l}\right)}\quad\mbox{and}\\
&a C_n^2(a)u^2+\left(C_{n+1}(a)+a C_{n-1}^{2^l}(a)\right)^2 u
\stackrel{(\ref{eq:Z})}{=}a C_n^2(a)u^2+Z_n^2(a) u=a\left(C_n^2(a)\right)^{2^l}\enspace.
\end{align*}
Note that the latter equation is a trivial identity when $C_n(a)=0$,
i.e., when $P_a(y)$ has more than two zeros (see
Proposition~\ref{pr:P2dZero}). Now use the substitution $u=v
Z_n^2(a)/a C_n^2(a)$ to obtain $v^2+v=\left(a
C_n^{2^l+1}(a)/Z_n^2(a)\right)^2$ (obviously, $C_n(a)\neq 0$ when
$Z_n(a)\neq 0$). Observe that
\begin{equation}
 \label{eq:C2lspl}
\frac{a C_n^{2^l+1}(a)}{Z_n^2(a)}\stackrel{(\ref{eq:C2l})}{=}
\frac{a C_{n-1}^{2^l}(a)C_{n+1}(a)}{Z_n^2(a)}+\frac{{\rm N}_d^k(a)}{Z_n^2(a)}
\stackrel{(\ref{eq:Z})}{=}\frac{C_{n+1}(a)}{Z_n(a)}+
\frac{C_{n+1}^2(a)}{Z_n^2(a)}+\frac{{\rm N}_d^k(a)}{Z_n^2(a)}
\end{equation}
and thus, ${\rm Tr}_k\left(a C_n^{2^l+1}(a)/Z_n^2(a)\right)=n{\rm
Tr}_d\left({\rm N}_d^k(a)/Z_n^2(a)\right)$ since
$Z_n(a)\in\mathrm{GF}(2^d)$. Therefore, if $n$ is odd and
$a^{2^l}x^{2^l+1}+x+a=0$ has exactly two roots in $\mathrm{GF}(2^k)$
then $Z_n(a)\neq 0$ and ${\rm Tr}_d\left({\rm
N}_d^k(a)/Z_n^2(a)\right)=0$.

For the case when $n$ is even, some additional arguments are needed.
Note that
\begin{eqnarray*}
\left(\frac{{\rm N}_d^k(a)}{Z_n^2(a)}\right)^2&\stackrel{(\ref{eq:C2lspl})}{=}&
\left(v+\frac{C_{n+1}^2(a)}{Z_n^2(a)}\right)+\left(v+\frac{C_{n+1}^2(a)}{Z_n^2(a)}\right)^2\quad\mbox{and}\\
\left(v+\frac{C_{n+1}^2(a)}{Z_n^2(a)}\right)^{2^l}&=&
\left(\frac{a C_n^{2^l+1}(a)}{Z_n^2(a)}\right)^2 v^{-1}+\frac{(C_n^2(a))^{2^l}+(C_{n+1}^2(a))^{2^l}}{Z_n^2(a)}\\
&\stackrel{(\ref{eq:dC1})}{=}&v+1+\frac{a^2(C_{n-1}^2(a))^{2^l}}{Z_n^2(a)}
\stackrel{(\ref{eq:Z})}{=}v+\frac{C_{n+1}^2(a)}{Z_n^2(a)}
\end{eqnarray*}
since $a^{2^l}u^{2^l+1}+u+a=0$ and using the relation between $u$
and $v$. Thus, $v+\frac{C_{n+1}^2(a)}{Z_n^2(a)}\in\mathrm{GF}(2^d)$
and ${\rm Tr}_d\left({\rm N}_d^k(a)/Z_n^2(a)\right)=0$.

Now prove the converse implication. Take an arbitrary $n$ and assume
$Z_n(a)\neq 0$ and ${\rm Tr}_d\left({\rm
N}_d^k(a)/Z_n^2(a)\right)=0$. Since ${\rm Tr}_k\left(a
C_n^{2^l+1}(a)/Z_n^2(a)\right)=0$ there exists some
$v\in\mathrm{GF}(2^k)$ with $v^2+v=\left(a
C_n^{2^l+1}(a)/Z_n^2(a)\right)^2$. Using the substitution $u=v
Z_n^2(a)/a C_n^2(a)$ we also obtain $a C_n^2(a)u^2+Z_n^2(a)
u=a(C_n^2(a))^{2^l}$. It is easy to see that
\begin{eqnarray*}
v^{2^l}+v&=&\sum_{i=1}^l\left(\frac{a
C_n^{2^l+1}(a)}{Z_n^2(a)}\right)^{2^i}\\
&\stackrel{(\ref{eq:C2lspl})}{=}&\frac{C_{n+1}^2(a)}{Z_n^2(a)}
+\frac{(C_{n+1}^{2^l}(a))^2}{Z_n^2(a)}+\frac{l}{d}{\rm Tr}_d\left(\frac{{\rm N}_d^k(a)}{Z_n^2(a)}\right)\\
&\stackrel{(\ref{eq:dC1},\ref{eq:Z})}{=}&\frac{(C_n^{2^l}(a)+Z_n(a))^2}{Z_n^2(a)}=\frac{(C_n^{2^l}(a))^2}{Z_n^2(a)}+1\enspace.
\end{eqnarray*}
Note that $n$ and $l/d$ can not be even together. Using the
substitution $u=v Z_n^2(a)/a C_n^2(a)$ we obtain
\begin{eqnarray*}
\left(a C_n^2(a)\right)^{2^l}u^{2^l}+a C_n^2(a)u&=&\left(C_n^{2^l}(a)\right)^2+Z_n^2(a)\quad\mbox{and}\\
\left(a C_n^2(a)\right)^{2^l}u^{2^l+1}+Z_n^2(a)u+a\left(C_n^2(a)\right)^{2^l}&=&u\left(\left(C_n^{2^l}(a)\right)^2+Z_n^2(a)\right)
\end{eqnarray*}
which gives $a^{2^l}u^{2^l+1}+u=a$. Thus, $P_a(x)$ has a zero and,
by Proposition~\ref{pr:P02Zero}, it has exactly two zeros.

Finally, note that the solution in $\mathrm{GF}(2^d)$ of the
equation $x^2+x=u$ for some $u\in\mathrm{GF}(2^d)$ with ${\rm
Tr}_d(u)=0$ and odd $d$ can be written as
$\sum_{i=0}^{\frac{d-1}{2}} u^{2^{2i}}$ or
$\sum_{i=0}^{\frac{d-3}{2}} u^{2^{2i+1}}$. This way we obtain
$v=W^2+\mu$ thus, $u=(W^2+\mu)Z_n^2(a)/a C_n^2(a)$ and
$P_{a^2}((W^2+\mu)Z_n^2(a)/C_n^2(a))=0$ for $\mu\in\{0,1\}$. It is
also not difficult to check by the direct calculations that
$P_a((W+\mu)Z_n(a)/C_n(a))=0$ if ${\rm Tr}_d\left({\rm
N}_d^k(a)/Z_n^2(a)\right)=0$.\qed
\end{pf}

In the case when $d=1$, if $Z_n(a)\neq 0$ (i.e., $Z_n(a)=1$) then
${\rm Tr}_d\left({\rm N}_d^k(a)/Z_n^2(a)\right)=0$ only for $a=0$
and thus, $P_a(x)$ has two zeros in $\mathrm{GF}(2^k)$ only for
$a=0$. The next proposition follows from
Propositions~\ref{pr:P02Zero} and \ref{pr:P2dZero}. We provide this
proof yet, independently of previous statements, since its major
part contains the result needed for proving the fact from
Corollary~\ref{co:ZeZ} and for the sake of giving the complete
picture of the addressed problem.

\begin{proposition}
 \label{pr:P1Zero}
Take any $a\in\mathrm{GF}(2^k)^*$. Then polynomial $P_a(x)$ has
exactly one zero in $\mathrm{GF}(2^k)$ if and only if $Z_n(a)=0$ and
$C_n(a)\neq 0$. Moreover, this zero is equal to $\left(a
C_n^{2^l-1}(a)\right)^{2^{k-1}}$ and if $n$ is odd (resp. $n$ is
even) then
\[M_1=2^{k-d}-1\quad (\mbox{resp.}\ 2^{k-d})\enspace.\]
\end{proposition}

\begin{pf}
Note that without loss of generality, we can substitute $a$ with
$a^2$ in the claimed result. First, assume $Z_n(a)=0$ and
$C_n(a)\neq 0$ (equivalently, we can take $a^2$). Now we find the
number of zeros of $L_a(x)$ in $\mathrm{GF}(2^k)$. Note that
\begin{eqnarray*}
L_a(C_n(a))&=&a_1 C_n^{2^{2l}}(a)+C_n^{2^l}(a)+a_0 C_n(a)\\
&\stackrel{(\ref{eq:dC1})}{=}&a_1 C_{n-1}^{2^{2l}}(a)+a_1 a_0
C_{n-2}^{2^{2l}}(a)+C_n^{2^l}(a)+a_0 C_n(a)\\
&\stackrel{(\ref{eq:dC2})}{=}&C_{n+1}(a)+a_0 C_{n-1}^{2^l}(a)\\
&\stackrel{(\ref{eq:Z})}{=}&Z_n(a)=0\enspace.
\end{eqnarray*}
Therefore, if $C_n(a)\neq 0$ then $2^d$ distinct elements $\mu
C_n(a)\in\mathrm{GF}(2^k)$ for $\mu\in\mathrm{GF}(2^d)$ are also
zeros of $L_a(x)$ (since
$\mathrm{GF}(2^k)\bigcap\mathrm{GF}(2^l)=\mathrm{GF}(2^d)$).

It is not difficult to see that in our case, $L_a(x)$ can not have
more than $2^d$ zeros in $\mathrm{GF}(2^k)$. Indeed, consider matrix
$\mathcal{M}_n$ of the system of $n$ linear equations
(\ref{eq:syst}). Note that $\det\mathcal{M}_n=Z_n^2(a)=0$ and a
principal submatrix obtained by deleting the last column and the
last row from $\mathcal{M}_n$ is nonsingular with the determinant
$\Delta_a(1,n-2)=C_n^2(a)\neq 0$ (see (\ref{eq:Delta})). Therefore,
applying equivalent row transformations to $\mathcal{M}_n$ we can
obtain a matrix containing a nonsingular diagonal submatrix lying in
the first $n-1$ columns and rows. Thus, the equation given by one of
the first $n-1$ rows (take row $i\in\{1,\dots,n-1\}$ with $il\equiv
d\ (\bmod\;k)$) of this equivalent matrix is nonzero and has degree
$2^d$. We conclude that system (\ref{eq:syst}) can not have more
than $2^d$ solutions and the same holds for the equation $L_a(x)=0$.
Note that on the side, we have found a factor
$x^{2^d}+C_n^{2^d-1}(a)x$ of $L_a(x)$ that contains all its zeros in
$\mathrm{GF}(2^k)$.

By Note~\ref{no:1} and Theorem~\ref{th:Bl}~(\ref{it:2}),
$P_{a^2}(x)$ has exactly one zero in $\mathrm{GF}(2^k)$ that is
equal to $a(\mu C_n(a))^{2^l-1}=a C_n^{2^l-1}(a)$ and the identities
for $M_1$ follow as well.

Now we prove the converse implication. Assume $P_{a^2}(x)$ has
exactly one zero in $\mathrm{GF}(2^k)$. Here we use the technique
found by Bluher \cite{Bl04} for counting the number of
$b\in\mathrm{GF}(2^k)^*$ for which $f_b(y)=y^{2^l+1}+by+b$ has
exactly one zero in $\mathrm{GF}(2^k)$. For any $v\in S$ with $S$
coming from (\ref{eq:S}) define
$r=v^{1-2^l}+1\in\mathrm{GF}(2^k)\setminus\{0,1\}$ and corresponding
$b=\frac{r^{2^l+1}}{r+1}\neq 0$. Obviously, such an $r$ is a zero of
$f_b(y)$. Note that
\begin{equation}
 \label{eq:b1}
b=\frac{r^{2^l+1}}{r+1}=v^{2^l-1}(v^{1-2^l}+1)^{2^l+1}=
\frac{(v+v^{2^l})^{2^l+1}}{v^{2^{2l}+1}}=V^{-1}\enspace,
\end{equation}
where $V$ comes from (\ref{eq:V}). Then, by Proposition~\ref{pr:ZeC}
and Corollary~\ref{co:ZeZ}, $C_n(b^{-1})\neq 0$ and $Z_n(b^{-1})=0$.
By the implication already proved, $P_{b^{-1}}(x)$ has exactly one
zero in $\mathrm{GF}(2^k)$. After substituting $x$ in the latter
polynomial with $b^{-1} y$, we get polynomial $f_{b^{2^l}}(y)$
having the same number of zeros as $P_{b^{-1}}(x)$. Thus, $f_b(y)$
(as well as $f_{b^{2^l}}(y)$) has exactly one zero in
$\mathrm{GF}(2^k)$.

Now we prove that function (\ref{eq:b1}) that maps every $v\in S$ to
$b\in\mathrm{GF}(2^k)^*$ is a $(2^d-1)$-to-$1$ mapping. First, note
that $(2^l-1)$-power is a $(2^d-1)$-to-$1$ mapping of $S$ to
$\mathrm{GF}(2^k)^*$. Indeed, if $x\in S$ and $x^{2^l-1}=t$ then the
latter identity holds for all distinct $\delta x\in S$ with
$\delta\in\mathrm{GF}(2^d)^*$ since ${\rm Tr}^k_d(\delta
x)=\delta{\rm Tr}^k_d(x)\neq 0$ and $\delta
x\notin\mathrm{GF}(2^d)$. Thus, every $r=v^{1-2^l}+1$ is obtained
from $2^d-1$ different values of $v$. Finally, the mapping from $r$
to $b$ is $1$-to-$1$ since for the obtained $b$ the equation
$f_b(y)=0$ has exactly one root $r$.

Therefore, taking all $v\in S$ and using (\ref{eq:b1}), we obtain
$|S|/(2^d-1)$ different values of $b\in\mathrm{GF}(2^k)^*$ and this
number is equal to the total number of $b$ such that $f_b(y)$ has
exactly one zero (see Theorem~\ref{th:Bl}). Therefore, these and
only these values of $b$ satisfying (\ref{eq:b1}) result in the
polynomials $f_b(y)$ having exactly one zero.

After substituting $x$ in $P_{a^2}(x)$ with $a^2 y$, we get
polynomial $f_b(y)$ with $b=a^{-2^{l+1}}$ having the same number of
zeros as $P_{a^2}(x)$. If polynomial $P_{a^2}(x)$ has exactly one
zero then $f_b(y)$ also has exactly one zero and, thus, $b$ is
obtained by (\ref{eq:b1}). Therefore, by Proposition~\ref{pr:ZeC},
$C_n(b^{-1})=C_n(a^{2^{l+1}})=C_n(a^2)^{2^l}\neq 0$ and, thus,
$C_n(a^2)\neq 0$. Also, by Corollary~\ref{co:ZeZ}, $Z_n(a^2)=0$.\qed
\end{pf}

Note that if $Z_n(a)=0$ and $C_n(a)\neq 0$ then, by
Proposition~\ref{pr:ZeC} and Corollary~\ref{co:ZeZ}, $a$ has the
form of (\ref{eq:V}) for some
$v\in\mathrm{GF}(2^k)\setminus\mathrm{GF}(2^d)$ with ${\rm
Tr}^k_d(v)\neq 0$ and
\begin{eqnarray*}
a C_n^{2^l+1}(a)&=&{\rm Tr}^k_d(v)^2{\rm N}^k_d\left(\frac{v}{v+v^{2^l}}\right)^2
v^{-2^{l+1}}\quad\quad\mbox{so}\\
{\rm Tr}^k_d\left(a^{-1}C_n^{-(2^l+1)}(a)\right)
&=&{\rm N}^k_d\left(1+v^{2^l-1}\right)^2\neq 0\enspace.
\end{eqnarray*}
This complies with the trace property from
Theorem~\ref{th:Bl}~(\ref{it:2}). Indeed, take $b=a_1^{-1}$ and
$\delta=\mu a C_n(a)$ for any $\mu\in\mathrm{GF}(2^d)^*$ then
$g(\delta)=\frac{L_a(C_n(a))}{a^{2^l+1}C_n(a)}=0$ and
$b^{-1}\delta^{-(2^l+1)}=\mu^{-2}a^{-1}C_n^{-(2^l+1)}(a)$.

Now we are left with the remaining case when $C_n(a)=0$ (then, by
Corollary~\ref{co:ZeZ}, $Z_n(a)=0$). The next proposition follows
from Propositions~\ref{pr:P02Zero} and \ref{pr:P1Zero}. We provide
this proof yet, independently of previous statements, since its
major part contains the result used for proving the converse
implication and also needed for proving the fact from
Proposition~\ref{pr:ZeC}. It is also worth mentioning
\cite[Lemma~22]{DoFeHeRo06}, where the authors found an interesting
parametrization for the set containing all $2^d+1$ zeros of
$x^{2^l+1}+ax^{2^l}+bx+c$ in $\mathrm{GF}(2^k)$. The latter
polynomial is directly related to $P_a(x)$, as noted in the
introduction.

\begin{proposition}
 \label{pr:P2dZero}
Take any $a\in\mathrm{GF}(2^k)^*$. Then polynomial $P_a(x)$ has
exactly $2^d+1$ zeros in $\mathrm{GF}(2^k)$ if and only if
$C_n(a)=0$. Also if $n$ is odd (resp. $n$ is even) then
\[M_{2^d+1}=\frac{2^{k-d}-1}{2^{2d}-1}\quad(\mbox{resp.}\ \frac{2^{k-d}-2^d}{2^{2d}-1})\enspace.\]
\end{proposition}

\begin{pf}
Here we use the technique found by Bluher \cite{Bl04} for counting
the number of $b\in\mathrm{GF}(2^k)^*$ for which
$f_b(y)=y^{2^l+1}+by+b$ has $2^d+1$ zeros in $\mathrm{GF}(2^k)$.
Denote $G=\mathrm{GF}(2^k)\setminus\mathrm{GF}(2^{2d})$ and observe
that
\[\mathrm{GF}(2^k)\cap\mathrm{GF}(2^{2l})=
\mathrm{GF}(2^{d\gcd(n,2)})\subseteq\mathrm{GF}(2^{2d})\enspace.\]
Therefore, taking any $u\in\mathrm{GF}(2^k)$ such that
$u\notin\mathrm{GF}(2^{2d})$ implies $u^{2^{2l}}\neq u$ and
$(u+u^{2^l})^{2^l}\neq u+u^{2^l}$ or, equivalently,
$u+u^{2^l}\notin\mathrm{GF}(2^l)$ which is the same as
$u+u^{2^l}\notin\mathrm{GF}(2^d)$. Now we can define
$r=(u+u^{2^l})^{1-2^l}+1\in\mathrm{GF}(2^k)\setminus\{0,1\}$ and
corresponding $b=\frac{r^{2^l+1}}{r+1}\neq 0$. Obviously, such an
$r$ is a zero of $f_b(y)$. Define also $r_0=ru^{2^l-1}$ and
$r_1=r(u+1)^{2^l-1}$ and note that $r$, $r_0$ and $r_1$ are pairwise
distinct. Further,
\[f_b(r_0)=r^{2^l+1}u^{2^{2l}-1}+bru^{2^l-1}+b=b((r+1)u^{2^{2l}}+ru^{2^l}+u)/u=0\]
since $r(u+u^{2^l})^{2^l}=u+u^{2^{2l}}$ by the definition of $r$.
Also, similarly, we get
\[f_b(r_1)=\frac{b((r+1)(u+1)^{2^{2l}}+r(u+1)^{2^l}+(u+1))}{u+1}=\frac{b(r+1+r+1)}{u+1}=0\enspace.\]
Thus, $f_b(y)$ with such a $b$ has at least three zeros and, by
Theorem~\ref{th:Bl}, it has $2^d+1$ zeros. Note that
\begin{equation}
 \label{eq:b}
b=\frac{r^{2^l+1}}{r+1}=(u+u^{2^l})^{2^l-1}((u+u^{2^l})^{1-2^l}+1)^{2^l+1}=
\frac{(u+u^{2^{2l}})^{2^l+1}}{(u+u^{2^l})^{2^{2l}+1}}=V^{-1}\enspace,
\end{equation}
where $V$ comes from (\ref{eq:V}) assuming $v=u+u^{2^l}$.

Now we prove that function (\ref{eq:b}) that maps every $u\in G$ to
$b\in\mathrm{GF}(2^k)^*$ is a $(2^{3d}-2^d)$-to-$1$ mapping. First,
note that $u\mapsto u+u^{2^l}$ is a $2^d$-to-$1$ mapping onto
\[F=\{x\in\mathrm{GF}(2^k)\setminus\mathrm{GF}(2^d)\ |\ {\rm
Tr}^k_d(x)=0\}\] (see explanations in the proof of
Proposition~\ref{pr:ZeC}). Further, $(2^l-1)$-power is a
$(2^d-1)$-to-$1$ mapping of $F$ to $\mathrm{GF}(2^k)^*$. Indeed, if
$x\in F$ and $x^{2^l-1}=t$ then the latter identity holds for all
distinct $\delta x\in F$ with $\delta\in\mathrm{GF}(2^d)^*$ since
${\rm Tr}^k_d(\delta x)=\delta{\rm Tr}^k_d(x)=0$ and $\delta
x\notin\mathrm{GF}(2^d)$. Thus, every $r=(u+u^{2^l})^{1-2^l}+1$ is
obtained from $2^d(2^d-1)$ different values of $u$. Finally, the
mapping from $r$ to $b$ is $(2^d+1)$-to-$1$ since for the obtained
$b$ the equation $f_b(y)=0$ has $(2^d+1)$ roots and every root $r$
satisfies $(r+1)^{-1}\in F^{2^l-1}$. Indeed, let $r$, $r_0$ and
$r_1$ be any distinct zeros of $f_b(y)$ (not necessarily the ones
defined above) and define $u=(r+r_1)/(r_0+r_1)$. Note that
\[rr_0(r+r_0)^{2^l}=r_0 r^{2^l+1}+r r_0^{2^l+1}=r_0 b(r+1)+r b(r_0+1)=b(r+r_0)\]
and so $b=rr_0(r+r_0)^{2^l-1}=rr_1(r+r_1)^{2^l-1}=r_0
r_1(r_0+r_1)^{2^l-1}$. Then
\[u^{2^{2l}-1}=(r_0/r)^{2^l+1}=(r_0+1)/(r+1)\neq 1\]
and, thus, $u\in G$. The identity $(r+1)^{-1}=(u+u^{2^l})^{2^l-1}\in
F^{2^l-1}$ follows from \cite[Lemma~2.1]{Bl04}.

Therefore, taking all $u\in G$ and using (\ref{eq:b}), we obtain
$|G|/(2^{3d}-2^d)$ different values of $b\in\mathrm{GF}(2^k)^*$ and
this number is equal to the total number of $b$ such that $f_b(y)$
has $2^d+1$ zeros (see Theorem~\ref{th:Bl}). Therefore, these and
only these values of $b$ satisfying (\ref{eq:b}) result in the
polynomials $f_b(y)$ having $2^d+1$ zeros.

Now note that without loss of generality, we can put $a^2$ in place
of $a$ in the result we are claiming. Then, after substituting $x$
in $P_{a^2}(x)$ with $a^2 y$, we get polynomial $f_b(y)$ with
$b=a^{-2^{l+1}}$ having the same number of zeros as $P_{a^2}(x)$. In
particular, polynomial $P_{a^2}(x)$ has exactly $2^d+1$ zeros in
$\mathrm{GF}(2^k)$ if and only if the same holds for the
corresponding polynomial $f_b(y)$ and this is equivalent to $b$
having the form of (\ref{eq:b}). It remains to apply
Proposition~\ref{pr:ZeC} and note that
$C_n(b^{-1})=C_n(a^{2^{l+1}})=C_n^{2^l}(a^2)$ and the latter is
equal to zero if and only if $C_n(a^2)=0$. The identities for
$M_{2^d+1}$ follow from Note~\ref{no:1} and
Theorem~\ref{th:Bl}~(\ref{it:3}).\qed
\end{pf}

Note that if $Z_n(a)=0$ then, by Corollary~\ref{co:ZeZ}, $a$ has the
form of (\ref{eq:V}) for some
$v\in\mathrm{GF}(2^k)\setminus\mathrm{GF}(2^d)$ and it is
straightforward to check that $P_a\left(v/(v+v^{2^l})\right)=0$.
This also complies with Proposition~\ref{pr:P1Zero} since if,
additionally, $C_n(a)\neq 0$ then, by (\ref{eq:CiV}), $\left(a
C_n^{2^l-1}(a)\right)^{2^{k-1}}=v/(v+v^{2^l})$. The following
corollary follows by combining Theorem~\ref{th:Pa} and
Proposition~\ref{pr:P1Zero}.

\begin{corollary}
Take any $a\in\mathrm{GF}(2^k)^*$ and positive integer $l<k$ with
$\gcd(l,k)=1$. Then ${\rm Tr}_k(R(a^{-1})+1)=1$ if and only if
$Z_k(a)=0$ and $C_k(a)\neq 0$, where $R(x)$, $C_k(x)$ and $Z_k(x)$
are defined in (\ref{eq:R}), (\ref{eq:dC1}) and (\ref{eq:Z})
respectively.
\end{corollary}

\section{Related Affine Polynomial}
 \label{sec:RelA}
In this section, we consider zeros in $\mathrm{GF}(2^k)$ of the
affine polynomial $F_a(x)$ defined in (\ref{eq:F}). Obviously,
either $F_a(x)$ has no zeros in $\mathrm{GF}(2^k)$ or it has exactly
the same number of zeros as its linearized homogeneous part $L_a(x)$
defined in (\ref{eq:L}). It was shown in Note~\ref{no:1} that
$L_a(x)/x$ and polynomial $g(x)$ from Theorem~\ref{th:Bl} are
related by a one-to-one substitution of variable. On the other hand,
$P_a(x)$ and polynomial $f(x)$ from the same theorem are related in
a similar way. Thus, we can use Theorem~\ref{th:Bl} and the results
from the previous sections to analyze the number of zeros of
$F_a(x)$ if we show that is has at least one zero in
$\mathrm{GF}(2^k)$. Moreover, since $L_a(cv)=cL_a(v)$ for any
$v\in\mathrm{GF}(2^k)$ and $c\in\mathrm{GF}(2^d)$, where
$d=\gcd(l,k)$, we can equivalently assume
\[F_a(x)=a^{2^l}x^{2^{2l}}+x^{2^l}+ax+c\enspace.\]
We know already from Lemma~\ref{le:V} that if $d=1$ and $a\neq 0$
then $cR(a^{-1})$, where $R(x)$ comes from (\ref{eq:R}), is a zero
of $F_a(x)$. Recall the notation $n=k/d$ and let also
\[N_i=\{a\;|\;a\in\mathrm{GF}(2^k)^*\ \mbox{and}\ F_a(x)\ \mbox{has exactly $i$ zeros in}\ \mathrm{GF}(2^k)\}\enspace.\]

\begin{lemma}
 \label{le:RootTr}
Take any $a\in\mathrm{GF}(2^k)^*$ and assume $F_a(x)$ has a zero in
$\mathrm{GF}(2^k)$, say $F_a(\mathcal{V})=0$. Then for any
$v\in\mathrm{GF}(2^k)$ with $F_a(v)=0$ holds
\[{\rm Tr}_d^k(v)={\rm Tr}_d^k(\mathcal{V})\in\{0,c\}\enspace.\]
Moreover, if $d=1$ then
\[\begin{array}{llll}
{\rm Tr}_k(a v^{2^l+1})&=&l'{\rm Tr}_k(R(a^{-1}))+{\rm Tr}_k(l'+1),&\quad\mbox{if}\quad v=R(a^{-1})\\
&=&l'{\rm Tr}_k(R(a^{-1}))+{\rm Tr}_k(l'),&\quad\mbox{if}\quad v\neq
R(a^{-1})\enspace,
\end{array}\]
where $R(x)$ is defined in (\ref{eq:R}) and $l'=l^{-1}\ (\bmod\;k)$.
\end{lemma}

\begin{pf}
The first identity follows by observing that any zero of $F_a(x)$ is
obtained as a sum of $\mathcal{V}$ and a zero of its homogeneous
part $L_a(x)$. To prove the identity it therefore suffices to show
that ${\rm Tr}_d^k(u)=0$ for any $u$ with
$a^{2^{l}}u^{2^{2l}}+u^{2^l}+au=0$. This follows from
\[{\rm Tr}_d^k(u)^2={\rm Tr}_d^k(u^2)={\rm Tr}_d^k(u^{2^{l+1}})={\rm Tr}_d^k(u^{2^l+2^l})
={\rm Tr}_d^k(a^{2^l}u^{2^{2l}+2^l}+au^{2^l+1})=0\] and thus, ${\rm
Tr}_d^k(u)=0$ as claimed. Also similarly, we get
\[{\rm Tr}_d^k(\mathcal{V})^2={\rm Tr}_d^k(\mathcal{V}^{2^l+2^l})
={\rm
Tr}_d^k(a^{2^l}\mathcal{V}^{2^{2l}+2^l}+a\mathcal{V}^{2^l+1}+c\mathcal{V}^{2^l})
={\rm Tr}_d^k(c\mathcal{V}^{2^l})=c{\rm Tr}_d^k(\mathcal{V})\] which
holds if and only if ${\rm Tr}_d^k(\mathcal{V})\in\{0,c\}$.

Now assume $\gcd(l,k)=1$ and let $\mathcal{V}=R(a^{-1})$ which, by
Lemma~\ref{le:V}, is a zero of $F_a(x)$ in $\mathrm{GF}(2^k)$. To
prove the second identity for the case when $v=\mathcal{V}$, we use
the fact presented in the proof of Lemma~\ref{le:V} that
$a\mathcal{V}^{2^l+1}=\sum_{i=1}^{l'}\mathcal{V}^{2^{il}}+l'+1$.
Then ${\rm Tr}_k(a\mathcal{V}^{2^l+1})=l'{\rm
Tr}_k(\mathcal{V})+{\rm Tr}_k(l'+1)$.

Now note that since $F_a(v)$ is obtained by adding the $2^l$th power
of (\ref{eq:Dob}) to itself we have for $v\neq 0$
\[F_a(v)=0\quad\mbox{if and only if}\quad
av^{2^l+1}+\sum_{i=1}^{l'}v^{2^{il}}+l'+1\in\{0,1\}\enspace.\] Since
$\mathcal{V}$ is the only solution of (\ref{eq:Dob}), then for
$v\neq\mathcal{V}$ with $F_a(v)=0$ we have
$av^{2^l+1}+\sum_{i=1}^{l'}v^{2^{il}}+l'+1=1$ and
\[{\rm Tr}_k(a
v^{2^l+1})=l'{\rm Tr}_k(v)+{\rm Tr}_k(l')=l'{\rm
Tr}_k(\mathcal{V})+{\rm Tr}_k(l')\] using already proved identity
that ${\rm Tr}_k(v)={\rm Tr}_k(\mathcal{V})$.\qed
\end{pf}

\begin{proposition}
 \label{pr:F1Zero}
Take any $a\in\mathrm{GF}(2^k)$. Then polynomial $F_a(x)$ has
exactly one zero in $\mathrm{GF}(2^k)$ if and only if $Z_n(a)\neq
0$. Moreover, this zero is equal to $\mathcal{V}_a=c C_n(a)/Z_n(a)$
and ${\rm Tr}_d^k(\mathcal{V}_a)=nc$. Also if $n$ is odd (resp. $n$
is even) then
\[|N_1|=\frac{2^{k+2d}-2^{k+d}-2^k+1}{2^{2d}-1}\ (\mbox{resp.}\
\frac{2^{k+2d}-2^{k+d}-2^k-2^{2d}+2^d+1}{2^{2d}-1})\ .\]
\end{proposition}

\begin{pf}
Having Theorem~\ref{th:Bl} and Proposition~\ref{pr:P02Zero}, it
suffices to show that $\mathcal{V}_a$ indeed is a zero of $F_a(x)$
if $Z_n(a)\neq 0$. First, recall that $Z_n(u)\in\mathrm{GF}(2^d)$
for any $u\in\mathrm{GF}(2^k)$. Therefore,
\begin{eqnarray}
 \label{eq:F(C)}
F_a(\mathcal{V}_a)
&=&\frac{c}{Z_n(a)}\left(a_1 C_n^{2^{2l}}(a)+C_n^{2^l}(a)+a_0 C_n(a)+Z_n(a)\right)\\
\nonumber&\stackrel{(\ref{eq:dC1})}{=}&\frac{c}{Z_n(a)}\left(a_1
C_{n-1}^{2^{2l}}(a)+a_1 a_0
C_{n-2}^{2^{2l}}(a)+C_n^{2^l}(a)+a_0 C_n(a)+Z_n(a)\right)\\
\nonumber&\stackrel{(\ref{eq:dC2})}{=}&\frac{c}{Z_n(a)}\left(C_{n+1}(a)+a_0
C_{n-1}^{2^l}(a)+Z_n(a)\right)=0\enspace.
\end{eqnarray}

To prove the trace identity for $\mathcal{V}_a$ first note that for
any $u\in\mathrm{GF}(2^k)$
\begin{eqnarray*}
{\rm Tr}_d^k(C_n(u)+Z_n(u))&\stackrel{(\ref{eq:Z})}{=}&{\rm Tr}_d^k\left(C_n(u)+C_{n+1}(u)+u_0 C_{n-1}^{2^l}(u)\right)\\
\nonumber&\stackrel{(\ref{eq:dC1})}{=}&{\rm Tr}_d^k\left(C_n(u)+C_n(u)+u_{n-1} C_{n-1}(u)+u_0 C_{n-1}^{2^l}(u)\right)\\
\nonumber&=&{\rm Tr}_d^k\left(u_{n-1} C_{n-1}(u)+(u_{n-1}
C_{n-1}(u))^{2^l}\right)=0\enspace.
\end{eqnarray*}
Therefore, since $c$ and $Z_n(a)$ are both in $\mathrm{GF}(2^d)$,
then
\[{\rm Tr}_d^k(\mathcal{V}_a)=
{\rm Tr}_d^k\left(c+c\,\frac{C_n(a)+Z_n(a)}{Z_n(a)}\right)
=nc+\frac{c}{Z_n(a)}\;{\rm Tr}_d^k(C_n(a)+Z_n(a))=nc\enspace.\]
Finally, $|N_1|=T_0+T_2$ taken from Theorem~\ref{th:Bl}.\qed
\end{pf}

Note that if $Z_n(a)\neq 0$ then, by Theorem~\ref{th:Bl} and
Proposition~\ref{pr:P02Zero}, the linear operator $L_a(x)$ on
$\mathrm{GF}(2^k)$ has the kernel of dimension zero and, thus, the
number of elements in the image of $L_a$ is $2^k$. Thus, the
equation $L_a(x)=c$ has a solution for any $c\in\mathrm{GF}(2^k)$ if
$Z_n(a)\neq 0$. Also note that if $d=1$, $a\neq 0$ and $Z_k(a)\neq
0$, i.e., $Z_k(a)=1$, then, by Lemma~\ref{le:V}, $R(a^{-1})=C_k(a)$.

\begin{proposition}
 \label{pr:F2dZero}
Take any $a\in\mathrm{GF}(2^k)^*$. Then polynomial $F_a(x)$ has
exactly $2^d$ zeros in $\mathrm{GF}(2^k)$ if and only if $Z_n(a)=0$
and $C_n(a)\neq 0$. In this case, ${\rm Tr}_d^k(v)=(n-1)c$ and, if
$n$ is even, then ${\rm Tr}_k\left(a c^{-2} v^{2^l+1}\right)$ is
constant for any $v\in\mathrm{GF}(2^k)$ with $F_a(v)=0$. Moreover, if
$n$ is odd then these zeros are
\[v_\mu=c\sum_{i=0}^{\frac{n-1}{2}}\frac{C_{n-1}^{2^{(2i+1)l}}(a)}{C_n^{2^{(2i+1)l}+2^{2il}-1}(a)}+\mu C_n(a)\]
for every $\mu\in\mathrm{GF}(2^d)$ and
\[\sum_{\mu\in\mathrm{GF}(2^d)}(-1)^{{\rm Tr}_k\left(a c^{-2} v_\mu^{2^l+1}\right)}=0\enspace.\]
Also if $n$ is odd (resp. $n$ is even) then $|N_{2^d}|=2^{k-d}-1$
(resp. $2^{k-d}$).
\end{proposition}

\begin{pf}
Having Theorem~\ref{th:Bl} and Proposition~\ref{pr:P1Zero}, it
suffices to show that $F_a(x)$ has at least one zero in
$\mathrm{GF}(2^k)$ if $Z_n(a)=0$ and $C_n(a)\neq 0$. From now on
assume $Z_n(a)=0$, $C_n(a)\neq 0$. Note that in this case, by
(\ref{eq:F(C)}),
\begin{equation}
 \label{eq:Bhom}
a_1 C_n^{2^{2l}}(a)+C_n^{2^l}(a)+a_0 C_n(a)=0
\end{equation}
which means that all $2^d$ distinct elements $\mu
C_n(a)\in\mathrm{GF}(2^k)$ for $\mu\in\mathrm{GF}(2^d)$ are zeros of
$L_a(x)$, since
$\mathrm{GF}(2^k)\bigcap\mathrm{GF}(2^l)=\mathrm{GF}(2^d)$.

Consider the following equation over $\mathrm{GF}(2^k)$
\begin{equation}
 \label{eq:2d}
C_n(a)x^{2^l}+C_n^{2^l}(a)x=cC_{n-1}^{2^l}(a)\enspace.
\end{equation}
Substituting $x=C_n(a)y$ we obtain
\[y^{2^l}+y=\frac{cC_{n-1}^{2^l}(a)}{C_n^{2^l+1}(a)}\] which has a
solution since, by Corollary~\ref{co:ZeTr}, ${\rm
Tr}_d^k\left(\frac{C_{n-1}^{2^l}(a)}{C_n^{2^l+1}(a)}\right)=0$ (if
$n>2$) and $c\in\mathrm{GF}(2^d)$ (see explanations in the proof of
Proposition~\ref{pr:ZeC}).

Therefore, there exists some $u\in\mathrm{GF}(2^k)$ with
\[\begin{array}{rrrrl}
C_n(a)u_1&+&C_n^{2^l}(a)u_0&=&cC_{n-1}^{2^l}(a)\quad\mbox{and}\\
C_n^{2^l}(a)u_2+C_n^{2^{2l}}(a)u_1&&&=&cC_{n-1}^{2^{2l}}(a)\enspace,
\end{array}\]
where the second identity is obtained by raising the first one to
the power of $2^l$. Now, multiply the first identity by $a_0
C_n^{-2^l}(a)$, the second by $a_1 C_n^{-2^l}(a)$ and add them to
obtain
\begin{align*}
&a_1 u_2+C_n^{-2^l}(a)\left(a_0 C_n(a)+a_1 C_n^{2^{2l}}(a)\right)u_1+a_0 u_0\\
&\stackrel{(\ref{eq:Bhom})}{=}a_1 u_2+u_1+a_0 u_0\\
&=cC_n^{-2^l}(a)\left(a_0 C_{n-1}^{2^l}(a)+a_1 C_{n-1}^{2^{2l}}(a)\right)\\
&\stackrel{(\ref{eq:Z})}{=}cC_n^{-2^l}(a)\left(a_0 C_{n-1}^{2^l}(a)+C_{n+1}^{2^l}(a)\right)\stackrel{(\ref{eq:dC1})}{=}c\enspace.
\end{align*}
Thus, $F_a(u)=0$.

If $Z_n(a)=0$ and $C_n(a)\neq 0$ then, by Proposition~\ref{pr:ZeC}
and Corollary~\ref{co:ZeZ}, $a=v_0^{2^{2l}+1}/(v_0+v_1)^{2^l+1}$ for
some $v\in\mathrm{GF}(2^k)\setminus\mathrm{GF}(2^d)$ with ${\rm
Tr}^k_d(v_0)\neq 0$. If $u\in\mathrm{GF}(2^k)$ is a solution of
(\ref{eq:2d}) then $F_a(u)=0$ and, by (\ref{eq:CiV}),
\[\frac{v_2}{v_1+v_2}u_1+\frac{v_0}{v_0+v_1}u_0=c\frac{v_2+\cdots+v_n}{{\rm
Tr}^k_d(v_0)}\enspace.\] Taking the trace of the both sides, we get
\[{\rm
Tr}^k_d\left(\frac{v_1}{v_0+v_1}u_0+\frac{v_0}{v_0+v_1}u_0\right)={\rm
Tr}^k_d(u)=(n-1)c\enspace.\] By Lemma~\ref{le:RootTr}, all zeros of
$F_a(x)$ have the same trace in $\mathrm{GF}(2^d)$.

To prove the properties of ${\rm Tr}_k\left(a c^{-2}
v^{2^l+1}\right)$, where $F_a(v)=0$, we use the technique suggested
by Dobbertin for proving \cite[Theorem~1]{Do99} (we did this already
in the proof of Lemma~\ref{le:Trv0}). Assume $c=1$. If $F_a(x)$ has
$2^d$ zeros then, by Theorem~\ref{th:Bl}, $P_a(x)$ has exactly one
zero, say $x_0\in\mathrm{GF}(2^k)$ with $x_0^{2^l+1}+x_0=a$. Define
polynomial $Q(x)=ax^{2^l}+x_0^2 x+x_0$ and denote
$\Gamma=x_0^{2^l-1}+x_0^{-1}$ (obviously $\Gamma\neq 0$ since
$x_0\neq 1$). As in Lemma~\ref{le:Trv0}, we can write
$F_a(x)=Q(x)\left(Q(x)^{2^l-1}+\Gamma\right)$.

Since $P_a(x_0)=0$ and $x_0$ is the only zero of $P_a(x)$, polynomial
\[P_a(x+x_0)=(x+x_0)^{2^l+1}+(x+x_0)+a=x^{2^l+1}+x_0 x^{2^l}+x_0^{2^l}x+x\]
has none zeros in $\mathrm{GF}(2^k)^*$. Multiplying the latter
polynomial by $x_0/x$, we obtain that $x_0 x^{2^l}+x_0^2 x^{2^l-1}+a$
has none zeros in $\mathrm{GF}(2^k)$ and the same can be said about
its reciprocal that is equal to $Q(x)$. Thus, all $2^d$ zeros of
$F_a(x)$ are exactly zeros of $Q(x)^{2^l-1}+\Gamma$ and there exists
a {\em unique} $\Delta\in\mathrm{GF}(2^k)^*$ with
$\Delta^{2^l-1}=\Gamma^{-1}$ such that $Q(x)=\Delta^{-1}$ has $2^d$
solutions in $\mathrm{GF}(2^k)$ (note that $Q(x)+b$ has none, one or
$2^d$ zeros depending on $b\in\mathrm{GF}(2^k)$). Also,
\[Q(\Delta)+x_0=\Delta\left(a\Delta^{2^l-1}+x_0^2\right)=
\Delta\left(\frac{x_0^{2^l+1}+x_0}{x_0^{2^l-1}+x_0^{-1}}+x_0^2\right)=0\]
and, therefore, taking a particular solution $v$ of
$Q(x)=\Delta^{-1}$, all the solutions are $v_\mu=v+\mu\Delta$ for
every $\mu\in\mathrm{GF}(2^d)$ (these $v_\mu$ are exactly the zeros
of $F_a(x)$). Now, using $Q(v_\mu)=\Delta^{-1}$, we obtain
\begin{eqnarray*}
{\rm Tr}_k\left(a v^{2^l+1}+a v_\mu^{2^l+1}\right)&=&
{\rm Tr}_k\left(x_0^2 v^2+x_0 v+\Delta^{-1}v+x_0^2 v_\mu^2+x_0 v_\mu+\Delta^{-1}v_\mu\right)\\
&=&{\rm Tr}_k\left(x_0^2\mu^2\Delta^2+x_0\mu\Delta+\mu\right)=n{\rm Tr}_d(\mu)
\end{eqnarray*}
and if $n$ is odd then
\[\sum_{\mu\in\mathrm{GF}(2^d)}(-1)^{{\rm Tr}_k\left(a v_\mu^{2^l+1}\right)}=
(-1)^{{\rm Tr}_k\left(a
v^{2^l+1}\right)}\sum_{\mu\in\mathrm{GF}(2^d)}(-1)^{{\rm
Tr}_d(\mu)}=0\enspace.\] This sum can also be calculated directly
(see \cite[Appendix~B]{HeKh08_2}). The case when $n$ is even comes
obviously. If $c\neq 1$, we need to multiply all zeros of $F_a(x)$ by
$c^{-1}$ to obtain zeros of $F_a(x)$ with $c=1$. Since
$c\in\mathrm{GF}(2^d)$ and $c^{2^l+1}=c^2$, we have the additional
coefficient $c^{-2}$ in the trace formulas.

Finally, note that the solution in $\mathrm{GF}(2^k)$ of the equation
$y^{2^l}+y=u$ for some $u\in\mathrm{GF}(2^k)$ with ${\rm
Tr}^k_d(u)=0$ and odd $n$ can be written in two ways as
$\sum_{i=0}^{\frac{n-1}{2}} u^{2^{2il}}$ or
$\sum_{i=0}^{\frac{n-3}{2}} u^{2^{(2i+1)l}}$ since ${\rm
Tr}_d^k(u)={\rm Tr}_l^{nl}(u)$ if $u\in\mathrm{GF}(2^k)$. It can also
be calculated directly that if $n$ is odd then $F_a(v_\mu)=0$ and
${\rm Tr}^k_d(v_\mu)=0$ (see \cite[Appendix~A]{HeKh08_2}). The
identities for $|N_{2^d}|$ follow from Theorem~\ref{th:Bl}.\qed
\end{pf}

\begin{proposition}
 \label{pr:F22dZero}
Take any $a\in\mathrm{GF}(2^k)$. Then polynomial $F_a(x)$ has at
least one zero in $\mathrm{GF}(2^k)$. Moreover, if $F_a(x)$ has
exactly $2^{2d}$ zeros then ${\rm Tr}_d^k(v)=nc$ for any
$v\in\mathrm{GF}(2^k)$ with $F_a(v)=0$ and, if $n$ is odd, then
\[\sum_{v\in\mathrm{GF}(2^k),\,F_a(v)=0}(-1)^{{\rm Tr}_k\left(a c^{-2} v^{2^l+1}\right)}=2^d\enspace.\]
Also if $n$ is odd (resp. $n$ is even) then
\[|N_{2^{2d}}|=\frac{2^{k-d}-1}{2^{2d}-1}\ (\mbox{resp.}\ \frac{2^{k-d}-2^d}{2^{2d}-1})\ .\]
\end{proposition}

\begin{pf}
Since the statement is obvious for $c=0$, we take $c\neq 0$. As
noted above, without loss of generality, we can also assume $c=1$.
Now, select any $u\in\mathrm{GF}(2^k)$ with ${\rm Tr}_d^k(u)=1$ and
fix it. Since ${\rm Tr}_d^k\left((u+u^2)^{2^l}\right)=0$, there
exists some $w\in\mathrm{GF}(2^k)$ such that
$w+w^{2^l}=(u+u^2)^{2^l}$ (see explanations in the proof of
Proposition~\ref{pr:ZeC}). Fix some $w$ with this property as well.

For any pair $(a,v)\in\mathrm{GF}(2^k)\times\mathrm{GF}(2^k)$ with
$F_a(v)=0$ and $v\neq u$, assuming
$b=\frac{av^{2^l+1}+w}{(v+u)^{2^l+1}}$, we have
\begin{eqnarray*}
&&F_b(v+u)=\frac{a^{2^l}v^{2^{2l}+2^l}+w^{2^l}}{(v+u)^{2^l}}+(v+u)^{2^l}
+\frac{av^{2^l+1}+w}{(v+u)^{2^l}}+1\\
&&\quad=\frac{1}{(v+u)^{2^l}}\left(v^{2^l}(a^{2^l}v^{2^{2l}}+v^{2^l}+av+1)
+w+w^{2^l}+(u+u^2)^{2^l}\right)=0\enspace.
\end{eqnarray*}
By Lemma~\ref{le:RootTr}, we obtain a $1$-to-$1$ correspondence
between two sets
\begin{eqnarray*}
S_0&=&\{(a,v)\;|\;v\neq u, F_a(v)=0, {\rm Tr}_d^k(v)=0\}\quad\mbox{and}\\
S_1&=&\{(a,v)\;|\;v\neq u, F_a(v)=0, {\rm Tr}_d^k(v)=1\}
\end{eqnarray*}
defined by $(a,v)\mapsto(b,v+u)$ and thus, $|S_0|=|S_1|$. Note that
for $n$ odd we can take $u=1$ and $w=0$.

Consider equation $F_x(u)=x^{2^l}u^{2^{2l}}+xu+u^{2^l}+1=0$ of the
unknown $x\in\mathrm{GF}(2^k)$. After substituting
$x=(uu^{2^l})^{-1}y$ we obtain equivalent equation
$y^{2^l}+y=(u+u^2)^{2^l}$ which has a solution in $\mathrm{GF}(2^k)$
since ${\rm Tr}_d^k\left((u+u^2)^{2^l}\right)=0$. Thus, $F_x(u)=0$
has exactly $2^d$ roots in $\mathrm{GF}(2^k)$ since its linearized
homogeneous part $x^{2^l}u^{2^{2l}}+xu$ has $2^d$ zeros which are
$\mu u^{-(2^l+1)}$ for every $\mu\in\mathrm{GF}(2^d)$.

Now, since $F_a(x)$ can have $0$, $1$, $2^d$ or $2^{2d}$ zeros,
using Propositions~\ref{pr:F1Zero} and \ref{pr:F2dZero} and
Lemma~\ref{le:RootTr}, we can compute the following sum in two
different ways
\begin{eqnarray*}
&&\sum_{(a,v):\,F_a(v)=0}(-1)^{{\rm Tr}_d^k(v)}=|S_0|-|S_1|-2^d\\
&&\quad\quad\quad=(-1)^{{\rm Tr}_d^k(1)}+\sum_{a\in N_1}(-1)^{{\rm Tr}_d^k(\mathcal{V}_a)}
+\sum_{a\in N_{2^d}}\sum_{v:\,F_a(v)=0}(-1)^{{\rm Tr}_d^k(v)}+X\\
&&\quad\quad\quad=(-1)^n\left(1+|N_1|-2^d|N_{2^d}|\right)+X=-2^d\enspace,
\end{eqnarray*}
where $X=\sum_{a\in N_{2^{2d}}}\sum_{v:\,F_a(v)=0}(-1)^{{\rm
Tr}_d^k(v)}$. Thus, if $n$ is odd (resp. $n$ is even) then
$X=-\frac{2^{2d}(2^{k-d}-1)}{2^{2d}-1}$ (resp.
$\frac{2^{2d}(2^{k-d}-2^d)}{2^{2d}-1}$) (note that $k=nd\geq 2d$ if
$n$ is even). Observe that the calculated values of $X$ satisfy
\[X=(-1)^n 2^{2d}(|\mathrm{GF}(2^k)^*|-|N_1|-|N_{2^d}|)\]
which, by Lemma~\ref{le:RootTr}, holds if and only only if ${\rm
Tr}_d^k(v)=n$ and $|N_{2^{2d}}|=\frac{2^{k-d}-1}{2^{2d}-1}$ (resp.
$\frac{2^{k-d}-2^d}{2^{2d}-1}$) for any $v\in\mathrm{GF}(2^k)$ with
$F_a(v)=0$ and $a\in N_{2^{2d}}$ if $n$ is odd (resp. $n$ is even).
Since $|N_1|+|N_{2^d}|+|N_{2^{2d}}|=|\mathrm{GF}(2^k)^*|$ and
$F_0(x)$ has a unique zero $x=1$, polynomial $F_a(x)$ has at least
one zero in $\mathrm{GF}(2^k)$ for any $a\in\mathrm{GF}(2^k)$.
Finally, note that zeros of $F_a(x)$ with an arbitrary
$c\in\mathrm{GF}(2^d)$ are exactly the elements $cv$ obtained from
every $v\in\mathrm{GF}(2^k)$ satisfying $F_a(v)=0$ with $c=1$
(obviously, ${\rm Tr}_d^k(cv)=nc$).

Now assume $n$ is odd and $c=1$. To prove the properties of ${\rm
Tr}_k\left(a c^{-2} v^{2^l+1}\right)$, where $F_a(v)=0$, we proceed
similarly to what we did in the proof of
Proposition~\ref{pr:F2dZero}. If $F_a(x)$ has $2^{2d}$ zeros then, by
Theorem~\ref{th:Bl}, $P_a(x)$ has $2^d+1$ zeros and we take one of
them, namely, $x_0\in\mathrm{GF}(2^k)$ with $x_0^{2^l+1}+x_0=a$.
Define polynomial $Q(x)=ax^{2^l}+x_0^2 x+x_0$ and denote
$\Gamma=x_0^{2^l-1}+x_0^{-1}$ (obviously $\Gamma\neq 0$ since
$x_0\neq 1$). As in Proposition~\ref{pr:F2dZero}, we can write
$F_a(x)=Q(x)\left(Q(x)^{2^l-1}+\Gamma\right)$.

Since $P_a(x_0)=0$ and $x_0$ is one of $2^d+1$ zeros of $P_a(x)$,
polynomial
\[P_a(x+x_0)=(x+x_0)^{2^l+1}+(x+x_0)+a=x^{2^l+1}+x_0 x^{2^l}+x_0^{2^l}x+x\]
has $2^d$ zeros in $\mathrm{GF}(2^k)^*$. Multiplying the latter
polynomial by $x_0/x$, we obtain that $x_0 x^{2^l}+x_0^2 x^{2^l-1}+a$
has $2^d$ zeros in $\mathrm{GF}(2^k)$ and the same can be said about
its reciprocal that is equal to $Q(x)$. Thus, $2^d$ zeros of $F_a(x)$
are also zeros of $Q(x)$ and the remaining $2^d(2^d-1)$ zeros of
$F_a(x)$ are also zeros of $Q(x)^{2^l-1}+\Gamma$. We conclude that
for every $\Delta\in\mathrm{GF}(2^k)^*$ with
$\Delta^{2^l-1}=\Gamma^{-1}$ (there are $2^d-1$ of such $\Delta$) we
can find $2^d$ solutions of $Q(x)=\Delta^{-1}$ in $\mathrm{GF}(2^k)$
(note that $Q(x)+b$ has none, one or $2^d$ zeros depending on
$b\in\mathrm{GF}(2^k)$).

For any $v\in\mathrm{GF}(2^k)$ with $Q(v)=0$ we have ${\rm
Tr}_k\left(a v^{2^l+1}\right)={\rm Tr}_k\left(x_0^2 v^2+x_0
v\right)=0$. Exactly in the same way as in
Proposition~\ref{pr:F2dZero}, we obtain that for odd $n$,
\[\sum_{v\in\mathrm{GF}(2^k),\,Q(v)=\Delta^{-1}}(-1)^{{\rm Tr}_k\left(a v^{2^l+1}\right)}=0\]
for any $\Delta\in\mathrm{GF}(2^k)^*$ with
$\Delta^{2^l-1}=\Gamma^{-1}$. In the general case when $c\neq 1$ we
need to multiply additionally the trace expression by $c^{-2}$.\qed
\end{pf}

Therefore, it can be concluded that polynomial $F_a(x)$ has exactly
$2^{2d}$ zeros in $\mathrm{GF}(2^k)$ if and only if $C_n(a)=0$.

\section{Related Linearized Polynomial}
 \label{sec:RelL}
In this section, we consider zeros in $\mathrm{GF}(2^{2k})$ of the
linearized polynomial
\begin{equation}
 \label{eq:Q}
Q_a(x)=r^{2^l}a^{2^l}x^{2^{2l}}+x^{2^{k+l}}+rax\enspace,
\end{equation}
where $l<k$, $a\in\mathrm{GF}(2^k)^*$ and
$r\in\mathrm{GF}(2^{2k})^*$ with $r^{2^k+1}=1$. For the details on
linearized polynomials in general, the reader is referred to Lidl
and Niederreiter \cite{LiNi97}. It is clear that $Q_a(x)$ does not
have multiple roots if $a\neq 0$ and that $Q_a(x)$ always has at
least one zero $x=0$.

Denote $d=\gcd(l,k)$, $n=k/d$ and $d_1=\gcd(k+l,2k)$. Observe that
\[Q_a(x)=(r^{-1} a)^{2^{k+l}}x^{2^{2(k+l)}}+x^{2^{k+l}}+rax\]
and in this form, polynomial $Q_a(x)$ reminds $L_a(x)$ defined in
(\ref{eq:L}). For any $u\in\mathrm{GF}(2^k)$ denote $u_i=u^{2^{il}}$
and let $r_i=r^{2^{il}}$ for $i=0,\dots,n-1$ so $Q_a(x)=r_1 a_1
x^{2^{2l}}+x^{2^{k+l}}+r_0 a_0 x$. Note that
$r^{2^{nl}}=r^{2^{kl/d}}=r^{(-1)^{l/d}}$. Dividing $Q_a(x)$ by $r_0
a_0 a_1^2 x$ (we remove one zero $x=0$) and using the substitution
$y=(ra)^{-1}x^{2^{k+l}-1}$ we obtain
\begin{equation}
 \label{eq:f}
f(y)=y^{2^{k+l}+1}+a_1^{-2}y+a_1^{-2}\enspace,
\end{equation}
the polynomial that appeared in Theorem~\ref{th:Bl}. Note that
multiplying $f(y)$ by $a_0^2 a_1^2$ and using the substitution
$z=a^2 y$ we obtain $z^{2^{k+l}+1}+z+a^2$, the polynomial having the
same number of zeros as $f(y)$ and that can be analyzed using the
results from Sections~\ref{sec:zer1}~and~\ref{sec:zer2}.

\begin{lemma}
 \label{le:rpow}
Take any $r\in\mathrm{GF}(2^{2k})^*$ with $r^{2^k+1}=1$. Then $r$ is
always a $(2^{k+l}-1)$-th power in $\mathrm{GF}(2^{2k})$ unless
$(k+l)/d$ is even and $r^{\frac{2^k+1}{2^d+1}}\neq 1$.
\end{lemma}

\begin{pf}
Note that $(k+l)/d$ is even if and only if both $n$ and $l/d$ are
odd which is equivalent to $d_1=2d$. In all other cases, $d_1=d$.
Let $\xi$ be a primitive element of $\mathrm{GF}(2^{2k})$. Then
$r=\xi^{(2^k-1)i}$ for some $i\in\{1,\dots,2^k+1\}$ and $r$ is a
$(2^{k+l}-1)$-th power in $\mathrm{GF}(2^{2k})$ if and only if there
exists some $j\in\{1,\dots,\frac{2^{2k}-1}{2^{d_1}-1}\}$ with
$r=\xi^{(2^{d_1}-1)j}$ since $\gcd(2^{k+l}-1,2^{2k}-1)=2^{d_1}-1$.
Now note that if $d_1=d$ then for any $i\in\{1,\dots,2^k+1\}$ there
exists some $j$ with $(2^k-1)i\equiv(2^{d_1}-1)j\ (\bmod\;2^{2k}-1)$
since $\gcd(2^{d_1}-1,2^{2k}-1)=2^d-1$ divides $(2^k-1)$. If
$d_1=2d$ then the above equivalence is solvable for $j$ if and only
if $\gcd(2^{d_1}-1,2^{2k}-1)=2^{2d}-1$ divides $(2^k-1)i$. The
latter holds if and only if $2^d+1$ divides $i$ since
$\gcd(2^d+1,2^k-1)=1$. Thus $r=\xi^{(2^k-1)(2^d+1)t}$ for some
$t\in\{1,\dots,\frac{2^k+1}{2^d+1}\}$ which is equivalent to
$r^{\frac{2^k+1}{2^d+1}}=1$.\qed
\end{pf}

\begin{proposition}
 \label{pr:Q}
For any $a\in\mathrm{GF}(2^k)^*$, take polynomials $Q_a(x)$ and
$f(y)$ over $\mathrm{GF}(2^{2k})$ defined in (\ref{eq:Q}) and
(\ref{eq:f}) respectively. If $(k+l)/d$ is even and
$r^{\frac{2^k+1}{2^d+1}}=1$ then exactly one of the following holds
\renewcommand{\theenumi}{\roman{enumi}}
\renewcommand{\labelenumi}{(\theenumi)}
\begin{enumerate}
\item $f(y)$ has one zero in $\mathrm{GF}(2^{2k})$ and $Q_a(x)$
    has $2^{2d}$ zeros in $\mathrm{GF}(2^{2k})$;

\item $f(y)$ has two zeros in $\mathrm{GF}(2^{2k})$ and $Q_a(x)$
    has one zero in $\mathrm{GF}(2^{2k})$;

\item $f(y)$ has $2^{2d}+1$ zeros in $\mathrm{GF}(2^{2k})$ and
    $Q_a(x)$ has $2^{4d}$ zeros in $\mathrm{GF}(2^{2k})$;
\end{enumerate}
If $(k+l)/d$ is odd then either
\renewcommand{\theenumi}{\roman{enumi}}
\renewcommand{\labelenumi}{(\theenumi)}
\begin{enumerate}
\item $f(y)$ has $2^d+1$ zeros in $\mathrm{GF}(2^{2k})$ and
    $Q_a(x)$ has $2^{2d}$ zeros in $\mathrm{GF}(2^{2k})$ or;

\item $f(y)$ has none or two zeros in $\mathrm{GF}(2^{2k})$ and
    $Q_a(x)$ has one zero in $\mathrm{GF}(2^{2k})$.
\end{enumerate}
\end{proposition}

\begin{pf}
Recall that $f(y)$ is obtained from $Q_a(x)/x$ using the
substitution $y=(ra)^{-1}x^{2^{k+l}-1}$ (there is also the
multiplicative constant that does not affect the number of zeros).
By Theorem~\ref{th:Bl}, $f(y)$ has $0$, $1$, $2$ or $2^{d_1}+1$
zeros in $\mathrm{GF}(2^{2k})$. Since raising elements of
$\mathrm{GF}(2^{2k})$ to the power of $(2^{k+l}-1)$ is a
$(2^{d_1}-1)$-to-$1$ mapping, polynomial $Q_a(x)$ can not have more
than $2^{2d_1}$ zeros in $\mathrm{GF}(2^{2k})$. Also, since zeros of
$Q_a(x)$ in $\mathrm{GF}(2^{2k})$ form a vector space over
$\mathrm{GF}(2^{d_1})$ then $Q_a(x)$ has $1$, $2^{d_1}$ or $2^{2
d_1}$ zeros in $\mathrm{GF}(2^{2k})$. Note that $d_1=d$ unless
$(k+l)/d$ is even which gives $d_1=2d$.

Assume $(k+l)/d$ is even and $Z_n(a)\neq 0$ where $Z_n(x)$ comes
from (\ref{eq:Z}) (note that $2k/d_1=n$). In this case,
\[{\rm N}_{d_1}^{2k}(a)={\rm N}_{2d}^{2k}(a)={\rm
N}_d^k(a)\in\mathrm{GF}(2^d)\] since $\gcd(2d,k)=d$, and ${\rm
Tr}_{d_1}\left({\rm N}_{d_1}^{2k}(a)/Z_n^2(a)\right)=0$ since
$Z_n(a)\in\mathrm{GF}(2^d)$. Therefore, by
Propositions~\ref{pr:P02Zero}~and~\ref{pr:P02ZeroD}, $f(y)$ always
has a zero in $\mathrm{GF}(2^{2k})$.

Now assume $f(y)$ has exactly one zero in $\mathrm{GF}(2^{2k})$ when
$(k+l)/d$ is odd (note that $d_1=d$). By Theorem~\ref{th:Bl}, this
is equivalent to
\[G(y)=y a_1 f\left(a_1^{-1}y^{2^{k+l}-1}\right)=a_2^{-1}y^{2^{2(k+l)}}+a_1^{-2}
y^{2^{k+l}}+a_1^{-1}y\] having $2^d$ zeros in $\mathrm{GF}(2^{2k})$.
Then there exists some $\mathcal{V}\in\mathrm{GF}(2^{2k})^*$ with
$G(\mathcal{V})=0$ and all zeros of $G(y)$ are exactly
$\{\mu\mathcal{V}\;|\;\mu\in\mathrm{GF}(2^d)\}$. Note that
$G(\mathcal{V}^{2^k})=G(\mathcal{V})^{2^k}=0$ since
$a\in\mathrm{GF}(2^k)$ and, thus,
$\mathcal{V}^{2^k-1}\in\mathrm{GF}(2^d)$. Take $\xi$ being a
primitive element of $\mathrm{GF}(2^{2k})$ and assume
$\mathcal{V}=\xi^i$. Then $\mathcal{V}^{2^k-1}\in\mathrm{GF}(2^d)$
if and only if $2^{2k}-1$ divides $i(2^k-1)(2^d-1)$ which is
equivalent to $2^k+1$ divide $i(2^d-1)$ and, further, to $2^k+1$
divide $i$ since $\gcd(2^k+1,2^d-1)=1$. Therefore,
$\mathcal{V}\in\mathrm{GF}(2^k)^*$ and ${\rm
Tr}^{2k}_d\left(a_1\mathcal{V}^{-(2^{k+l}+1)}\right)=0$ which
contradicts to the trace condition from
Theorem~\ref{th:Bl}~(\ref{it:2}). Thus, $f(y)$ can not have exactly
one zero in $\mathrm{GF}(2^{2k})$ under these conditions.

Assume $f(y)$ has exactly one or $2^{d_1}+1$ zeros in
$\mathrm{GF}(2^{2k})$ and $u$ is one of them. Then, by
Theorem~\ref{th:Bl}, there exists some $v\in\mathrm{GF}(2^{2k})$
with $u=a_1^{-1}v^{2^{k+l}-1}$ and the corresponding $g(v)=0$. In
this case, equation $u=(ra)^{-1}x^{2^{k+l}-1}$ is solvable for $x$
if and only if $r$ is a $(2^{k+l}-1)$-th power in
$\mathrm{GF}(2^{2k})$, every solution is a zero of $Q_a(x)$ and all
zeros are obtained this way from some $u$. Thus, by
Lemma~\ref{le:rpow}, $Q_a(x)/x$ has respectively $2^{d_1}-1$ or
$2^{2d_1}-1$ zeros in $\mathrm{GF}(2^{2k})$ unless $(k+l)/d$ is even
and $r^{\frac{2^k+1}{2^d+1}}\neq 1$. In the remaining case,
$Q_a(x)/x$ has none zeros in $\mathrm{GF}(2^{2k})$. This also means
that in this case, $Q_a(x)$ can not have $2^{2 d_1}$ zeros in
$\mathrm{GF}(2^{2k})$ since this leads to $f(y)$ having $2^{d_1}+1$
zeros.

Assume $f(y)$ has exactly two zeros in $\mathrm{GF}(2^{2k})$ (let
$u$ is one of them) and consider the cases when $r$ is a
$(2^{k+l}-1)$-th power in $\mathrm{GF}(2^{2k})$. Then, by
Theorem~\ref{th:Bl} (\ref{it:1}), $ua_1=x^{2^{k+l}-1}$ is not
solvable for $x$ in $\mathrm{GF}(2^{2k})$. Therefore, $ura=ua_1
a^{-2^{k+l}+1}r$ is not a $(2^{k+l}-1)$-th power in
$\mathrm{GF}(2^{2k})$ and $Q_a(x)/x$ has none zeros in
$\mathrm{GF}(2^{2k})$.\qed
\end{pf}

If $(k+l)/d$ is even and $r^{\frac{2^k+1}{2^d+1}}\neq 1$ then
\renewcommand{\theenumi}{\roman{enumi}}
\renewcommand{\labelenumi}{(\theenumi)}
\begin{enumerate}
\item $f(y)$ has one or $2^{2d}+1$ zeros in
    $\mathrm{GF}(2^{2k})$ and $Q_a(x)$ has one zero in
    $\mathrm{GF}(2^{2k})$;

\item $f(y)$ has two zeros in $\mathrm{GF}(2^{2k})$, $Y_n(a)\neq
    0$ and $Q_a(x)$ has one zero in $\mathrm{GF}(2^{2k})$;

\item $f(y)$ has two zeros in $\mathrm{GF}(2^{2k})$, $Y_n(a)=0$
    and $Q_a(x)$ has $2^{2d}$ zeros in $\mathrm{GF}(2^{2k})$.
\end{enumerate}

Note that if $(k+l)/d$ is even, $r^{\frac{2^k+1}{2^d+1}}\neq 1$ and
$Z_n(a)=0$ (the latter, by Note~\ref{no:1} and
Propositions~\ref{pr:P1Zero} and \ref{pr:P2dZero}, is equivalent to
$f(y)$ having $1$ or $2^{d_1}+1$ zeros) then $Q_a(x)$ has one zero
in $\mathrm{GF}(2^{2k})$ (observe that $2k/d_1=n$).

For $0<j\leq i$ and $u\in\mathrm{GF}(2^k)$, let $D^{j,i}_u$ denote a
three-diagonal matrix of size $i-j+2$ that contains ones on the main
diagonal and with
\[D^{j,i}_u(t,t+1)=r^{(-1)^{j+t-1}}_{j+t}u_{j+t}\quad\mbox{and}\quad
D^{j,i}_u(t+1,t)=r^{(-1)^{j+t}}_{j+t}u_{j+t}\] for $t=0,\dots,i-j$,
where indices of $u$ are reduced modulo $n$, indices of $r$ are
reduced using the rule $r_{tn+i}=r_i^{(-1)^{tl/d}}$ ($i=0,\dots,n-1$
and $t\geq 0$), rows and columns of $D^{j,i}_u$ are numbered from
$0$ to $i-j+1$. The determinant of $D^{j,i}_u$, denoted as
$\Delta'_u(j,i)$, can be computed expanding by minors along the last
row to obtain
\[\Delta'_u(j,i)=\Delta'_u(j,i-1)+u_i^2\Delta'_u(j,i-2)\]
assuming $\Delta'_u(j,i)=1$ if $i-j\in\{-2,-1\}$. Comparing the
latter recursive identity with (\ref{eq:RecDelta}) it is easy to see
that
\begin{equation}
 \label{eq:DeltaPr}
\Delta'_u(j,i)=\Delta_u(j,i)\enspace.
\end{equation}

\begin{proposition}
 \label{pr:L1Zero}
Let $(k+l)/d$ be even and take any $a\in\mathrm{GF}(2^k)^*$. Then
$Q_a(x)=0$ has exactly one root in $\mathrm{GF}(2^{2k})$ that is
equal to zero if $Z^2_n(a)\neq{\rm N}^k_d(a)(\delta+\delta^{-1})$,
where $\delta=r^{\frac{2^k+1}{2^d+1}}\in\mathrm{GF}(2^{2d})$ is a
$(2^d+1)$-th root of unity over $\mathrm{GF}(2)$ and $Z_n(x)$ comes
from (\ref{eq:Z}).
\end{proposition}

\begin{pf}
Note that $\delta+\delta^{-1}\in\mathrm{GF}(2^d)$ and thus,
$Y_n(u)\in\mathrm{GF}(2^d)$ for any $u\in\mathrm{GF}(2^k)$ since
$Z_n(u)\in\mathrm{GF}(2^d)$.

Obviously, $Q_a(0)=0$ and we have to show that this is the only zero
of $Q_a(x)$ in $\mathrm{GF}(2^{2k})$ if $Y_n(a)\neq 0$. Taking
equation $Q_a(x)=0$ and all its $2^{2li}$ powers we obtain $n$
equations
\[Q_a^{2^{2il}}(x)=x^{2^{(2i+1)l+k}}+r_{2i+1}a_{2i+1}x^{2^{2l(i+1)}}+r_{2i}a_{2i}x^{2^{2li}}=0\quad\mbox{for}\quad
i=0,\dots,n-1\enspace,\] where indices of $a$ are reduced modulo $n$
and indices of $r$ are reduced using the rule
$r_{tn+i}=r_i^{(-1)^{tl/d}}$ ($i=0,\dots,n-1$ and $t\geq 0$). If
$x_{2i}$ $(i=0,\dots,n-1)$ are considered as independent variables
then matrix $\mathcal{M}_n$ of the obtained system of $n$ linear
equations with $n$ unknowns consists of three cyclic antidiagonals
and
 \setlength{\arraycolsep}{0.14em}
\begin{eqnarray*}
\mathcal{M}_n(i,(n-3)/2-i)&=&1\ ,\\
\mathcal{M}_n(i,n-i-1)&=&r_{2i}a_{2i}\ ,\\
\mathcal{M}_n(i,n-i-2)&=&r_{2i+1}a_{2i+1}\quad\mbox{for}\quad
i=0,\dots,n-1\enspace,
\end{eqnarray*}
 \setlength{\arraycolsep}{5pt}\noindent
where rows and columns of $\mathcal{M}_n(i,j)$ are numbered from $0$
to $n-1$ and all elements of $\mathcal{M}_n$ are indexed modulo $n$.

Now permute the columns and rows of $\mathcal{M}_n$ in the following
way. Decimate the rows as $i(n+1)/2$ and columns as
$(n-3)/2+i(n-1)/2$ modulo $n$ for $i=0,\dots,n-1$ (note that
$\gcd((n+1)/2,n)=\gcd((n-1)/2,n)=1$). Then the obtained matrix
$\mathcal{M}'_n$ is three-diagonal cyclic with
 \setlength{\arraycolsep}{0.14em}
\begin{eqnarray*}
\mathcal{M}'_n(i,i)&=&\mathcal{M}_n(i(n+1)/2,(n-3)/2+i(n-1)/2)=1\ ,\\
\mathcal{M}'_n(i,i-1)&=&\mathcal{M}_n(i(n+1)/2,(n-3)/2+(i-1)(n-1)/2)=r_{i(n+1)}a_{i(n+1)}\ ,\\
\mathcal{M}'_n(i,i+1)&=&\mathcal{M}_n(i(n+1)/2,(n-3)/2+(i+1)(n-1)/2)=r_{i(n+1)+1}a_{i(n+1)+1}
\end{eqnarray*}
 \setlength{\arraycolsep}{5pt}\noindent
for $i=0,\dots,n-1$ (indices of $r$ and $a$ are calculated modulo
$2n$) since
 \setlength{\arraycolsep}{0.0em}
\begin{align*}
&i(n+1)/2+(n-3)/2+i(n-1)/2=(n-3)/2+in\equiv (n-3)/2\ (\bmod\;n)\ ,\\
&i(n+1)/2+(n-3)/2+(i-1)(n-1)/2=-1+in\equiv n-1\ (\bmod\;n)\ ,\\
&i(n+1)/2+(n-3)/2+(i+1)(n-1)/2=n-2+in\equiv n-2\
(\bmod\;n)\enspace.
\end{align*}
 \setlength{\arraycolsep}{5pt}\noindent
Also note that $a_{i(n+1)}=a_i$ since $a\in\mathrm{GF}(2^{nk})$ and
$r_{i(n+1)}=r^{(-1)^i}_i$ since $r_{n+i}=r^{2^{ik}}_n=r^{-1}_i$ for
any $i\geq 0$. Then for $i=0,\dots,n-1$
\[\mathcal{M}'_n(i,i+1)=r^{(-1)^i}_{i+1} a_{i+1}\quad\mbox{and}
\quad\mathcal{M}'_n(i+1,i)=r^{(-1)^{i+1}}_{i+1} a_{i+1}\quad\mbox{so}\]
\begin{equation*}
\mathcal{M}'_n=\left(\begin{array}{cccccc}
1&r_1 a_1&0&\cdots&r_0 a_0\\
r^{-1}_1 a_1&\ddots&\ddots&\ddots&0\\
\vdots&\ddots&\ddots&\ddots&\vdots\\
0&&\ddots&1&r^{-1}_{n-1} a_{n-1}\\
r^{-1}_0 a_0&0&\cdots&r_{n-1} a_{n-1}&1
\end{array}\right)\enspace.
\end{equation*}
Note that a principal submatrix obtained by deleting the last column
and the last row from $\mathcal{M}'_n$ is exactly $D^{1,n-2}_a$.

We also have to apply the decimation $(n-3)/2+i(n-1)/2$ modulo $n$
for $i=0,\dots,n-1$ (used to permute the columns of $\mathcal{M}$) to
the vector of unknowns $(z_{2(n-1)},z_{2(n-2)},\dots,z_2,z_0)$. This
results in $\vec{z}=(z_{n+1},z_2,z_{n+3},\dots,\allowbreak
z_{n-1},z_0)^{\rm T}$, where the increment for the index of $z$ is
equal to $n-1$ starting from $0$ and going right to left (indices are
calculated modulo $2n$). Now, if $\vec{0}=(0,\dots,0)^{\rm T}$ then a
new system has the following matrix representation
\begin{equation}
 \label{eq:syst1}
\mathcal{M}'_n\vec{z}=\vec{0}\enspace.
\end{equation}
The determinant of $\mathcal{M}_n$ is equal to the determinant of
$\mathcal{M}'_n$ and can be computed expanding the latter by minors
along the last row. Doing this it is easy to see that
 \setlength{\arraycolsep}{0.14em}
\begin{eqnarray*}
\det\mathcal{M}'_n&=&\Delta'_a(1,n-2)+r_{n-1}a_{n-1}
\left(r^{-1}_{n-1}a_{n-1}\Delta'_a(1,n-3)+\prod_{i=0}^{n-2}r^{(-1)^i}_i a_i\right)\\
&&\quad\quad\quad\quad\quad\ \ +\:r^{-1}_0 a_0\left(r_0 a_0\Delta'_a(2,n-2)+\prod_{i=1}^{n-1}r^{(-1)^{i-1}}_i a_i\right)\\
&\stackrel{(\ref{eq:Delta},\ref{eq:Delta2l},\ref{eq:DeltaPr})}{=}&B_n^2(a)+a_{n-1}^2
B_{n-1}^2(a)+(a_0 B_{n-1}^{2^k}(a))^2+{\rm
N}^{nk}_k(a)(\delta+\delta^{-1})\\
&\stackrel{(\ref{eq:dC1},\ref{eq:Z})}{=}&Z^2_n(a)+{\rm
N}^{nk}_k(a)(\delta+\delta^{-1})=Y_n(a)\enspace.
\end{eqnarray*}
 \setlength{\arraycolsep}{5pt}\noindent
Thus, if $Y_n(a)\neq 0$ then (\ref{eq:syst1}) has only zero
solution. Now note that every $v\in\mathrm{GF}(2^{2nk})$ with
$Q_a(v)=0$ provides a solution to the system given by
$v_{2i}=v^{2^{2ik}}$ for $i=0,\dots,n-1$. Therefore, if $Y_n(a)\neq
0$ then $Q_a(z)$ has at most one zero.\qed
\end{pf}

\section{Conclusion}
We studied the polynomials $P_a(x)=x^{2^l+1}+x+a$ over
$\mathrm{GF}(2^k)$ with $l<k$ and proved some new criteria for the
number of zeros of $P_a(x)$ in $\mathrm{GF}(2^k)$. In particular, the
number of zeros and the trace of the value of the polynomial, due to
Dobbertin, in point $a^{-1}$ are related when $\gcd(l,k)=1$. In case
when there is a unique zero or exactly two zeros and $\gcd(l,k)$ is
odd, we provided explicit expressions for calculating these roots as
polynomials of $a$. We also found the distribution of the number of
zeros of $P_a(x)$. Finally, we studied the affine polynomial
$F_a(x)=a^{2^l}x^{2^{2l}}+x^{2^l}+ax+c$ with
$c\in\mathrm{GF}(2^{\,\gcd(l,k)})$, which was shown to be closely
related to $P_a(x)$. In many cases, we were able to provide explicit
expressions for calculating zeros of $F_a(x)$.


\end{document}